\documentclass{aastex62}
\turnoffedit

\graphicspath{{./}{figures/}}

\shorttitle{Photometric Biases in Modern Surveys}
\shortauthors{Portillo, Speagle, \& Finkbeiner}
\usepackage{amsmath}

\DeclareMathAlphabet\mathbfcal{OMS}{cmsy}{b}{n} 
\newcommand{\mean}{\boldsymbol{\mu}}  
\newcommand{\cov}{\mathbf{C}}  
\newcommand{\normvec}{\mathbf{Z}}  
\newcommand{\zeros}{\mathbf{0}}  
\newcommand{\T}{\textrm{T}}  
\newcommand{\diag}{\textrm{diag}}  
\newcommand{\btensor}{\mathbfcal{B}}  

\newcommand{\xdim}{n}  
\newcommand{\ydim}{m}  
\newcommand{\bflux}{\mathbf{b}}  
\newcommand{\flux}{\mathbf{f}}  
\newcommand{\psf}{\mathbf{p}}  
\newcommand{\ml}{\textrm{ML}}  
\newcommand{\bparams}{\boldsymbol{\beta}}  
\newcommand{\bdim}{k}  
\newcommand{\params}{\boldsymbol{\theta}}  
\newcommand{\param}{\theta}  
\newcommand{\fisher}{\mathbfcal{F}}  
\newcommand{\neff}{A_{\rm psf}}  
\newcommand{\seff}{S_{\rm psf}}  
\newcommand{\bias}{\delta}  

\begin{document}

\title{Photometric Biases in Modern Surveys}

\author[0000-0001-8132-8056]{Stephen K. N. Portillo}
\altaffiliation{Equal contribution}
\affil{DIRAC Institute, Department of Astronomy\\
University of Washington\\
3910 15th Ave NE\\
Seattle, WA 98195, USA}

\author[0000-0003-2573-9832]{Joshua S. Speagle}
\altaffiliation{Equal contribution}
\affil{Center for Astrophysics | Harvard \& Smithsonian \\
60 Garden St, MS-10 \\
Cambridge, MA 02138, USA}

\author[0000-0003-2808-275X]{Douglas P. Finkbeiner}
\affil{Center for Astrophysics | Harvard \& Smithsonian \\
60 Garden St, MS-10 \\
Cambridge, MA 02138, USA}

\correspondingauthor{Stephen K. N. Portillo}
\email{sportill@uw.edu, jspeagle@cfa.harvard.edu, dfinkbeiner@cfa.harvard.edu}

\begin{abstract}
\edit1{Many} surveys use maximum-likelihood (ML) methods to fit models when extracting photometry from images. We show these ML estimators systematically \textit{overestimate} the flux as a function of the signal-to-noise ratio and the number of model parameters involved in the fit. This bias is substantially worse for \edit1{resolved sources}: while a $1\%$ bias is expected for a $10\sigma$ point source, a $10\sigma$ \edit1{resolved} galaxy with a simplified Gaussian profile suffers a $2.5\%$ bias. This bias also behaves differently depending how multiple bands are used in the fit: simultaneously fitting all bands leads the flux bias to become roughly evenly distributed between them, while fixing the position in ``non-detection'' bands (i.e. forced photometry) gives flux estimates in those bands that are biased \textit{low}, compounding a bias in derived colors. We show that these effects are present in idealized simulations, outputs from the \edit1{Hyper Suprime-Cam} fake object pipeline (SynPipe), and observations from \edit1{Sloan Digital Sky Survey} Stripe 82. \edit1{Prescriptions to correct for the ML bias in flux, and its uncertainty, are provided.}
\end{abstract}


\section{Introduction}
\label{sec:intro}

\edit1{Precise flux density measurements are crucial for many areas of astronomy, from analyzing time-variable signals when searching for exoplanets \citep{Koch_2010} to determining precise colors and bolometric luminosities for stellar and galactic modeling \citep{2002Ap&SS.280....1P}.} For example, the \edit1{Large Synoptic Survey Telescope (LSST) will require photometric precision of $\sim 1\%$ for} weak-lensing studies, supernova cosmology, classifying potentially hazardous asteroids, and to separate out main sequence and giant stars to map the galaxy \edit1{\citep{2019ApJ...873..111I}}.  \edit1{Photometric accuracy is also important for measuring photometric redshifts \citep{2000A&A...363..476B}. By using multi-band forced photometry to improve the photometric accuracy, \cite{Nyland_2017} significantly improve photometric redshift accuracy in the Spitzer Extragalactic Representative Volume Survey. In a survey with billions of objects, statistical uncertainty (which generally averages down with the square root of the number of objects) may be generically less important than systematic bias (which does not).}

\edit1{Historically,} the gold standard for precision photometry was photoelectric measurement with a photomultiplier tube \citep[e.g.][]{Landolt1973,Landolt1983}. The photons entering an aperture of, e.g., an eight arcsecond radius were detected and counted. In this way, hundreds of stars could be measured with great precision, and these ``standard stars'' formed the basis of the commonly used photometric systems. \edit1{Landolt's repeat} measurements of these standards were impressively consistent, but even the ``Landolt Faint Standards'' \citep{Landolt1992} are relatively bright compared to the saturation limit of modern wide-area surveys.

Modern surveys use arrays of pixelated sensors such as charge-coupled devices (CCDs) in the optical/UV bands and mercury-cadmium-telluride (HgCdTe) devices in the near infrared. Surveys then perform ``aperture photometry'' by adding up all (background-subtracted) counts inside of a circle \edit1{or ellipse} in an image. This process works well for isolated stars well above the background noise, and forms the basis of, e.g., the Sloan Digital Sky Survey \edit1{\citep[SDSS;][]{2000AJ....120.1579Y}} photometric calibration \citep{Padmanabhan2008}. \edit1{Aperture photometry is widely used in astronomical surveys including the Two Micron All Sky Survey \citep{Skrutskie_2006}, the Cosmological Evolution Survey \citep[COSMOS;][]{2007ApJS..172....1S}, and the Transiting Exoplanet Survey Satellite \citep{2014SPIE.9143E..20R}}

Aperture photometry, however, has several significant issues that hamper its usage. \edit1{To place the aperture, the center of the point spread function (PSF) must be estimated with techniques like the image centroiding method, but these estimates may be inadequate if the source is crowded or faint \citep{1992ASPC...23...90D}.}
\edit1{To gather nearly all the light from a single source, apertures need to be relatively large compared to the seeing of the image. Since all pixels (except those on the boundary) in an aperture have equal statistical weight, including those with very little flux from the source, enlarging the aperture can reduce the signal-to-noise ratio (SNR) of faint sources closer to the background \citep{1989PASP..101..616H}.}
Shrinking the size of the aperture can \edit1{mitigate} this problem but then results in some amount of light from most sources being excluded, necessitating an ``aperture correction'' that needs to be calibrated \edit1{\citep{1989PASP..101..616H,1990PASP..102..932S}}.
Aperture photometry can also become quite sensitive to issues relating to background estimation: because all counts in the aperture are being added together, these systematic offsets contribute an increasing portion of the counts with increasing aperture size \edit1{\citep{1989PASP..101..616H}}.

\edit1{Motivated by the need to photometer crowded fields, astronomers introduced PSF photometry in algorithms like ROMAFOT \citep{1983A&A...126..278B} and DAOPHOT \citep{Stetson1987} which apply a 2-D matched filter \citep{Turin1960} to the image.} \edit1{For} point sources the appropriate 2-D ``filter'' is the point-spread function. Put another way, PSF photometry involves a parametric generative model of the noiseless data as a function of some parameters $\params$. Combining this with an appropriate noise model then yields a likelihood function. Maximizing the likelihood then provides an estimate of the flux with the highest signal-to-noise ratio. \edit1{For resolved sources, the appropriate filter is the PSF-convolved profile of the source: in model-fitting photometry, a parameterized model profile (eg. exponential, de Vaucouleurs) is used to approximate the source profile. Photometry then involves estimating the model profile parameters as well as the flux of the source.}

Even for point sources, this process is not always straightforward. While aperture photometry simply requires an appropriately-sized aperture, PSF photometry relies on having an accurate model of the underlying source. This requires knowing the PSF precisely across the image to avoid applying a mismatched filter. \edit1{The PSF is often extracted from the image itself using algorithms like PSFEx \citep{2013ascl.soft01001B}.} While for space-based telescopes the PSF can be quite stable, for ground-based surveys it varies as a function of time and position in the focal plane to such a degree that it can cause flux errors of $1-2\%$, dominating the error budget of bright stars \citep[e.g. ][]{Padmanabhan2008}.

Furthermore, while PSF photometry can perform better than aperture photometry for fainter objects by avoiding ``overweighting'' the background, it still struggles in crowded regions where the measured parameters for objects become covariant with those of close neighbors (or even bright neighbors that are not so close). In the extremely crowded limit, the challenge of estimating the PSF and the background level may become severe. These questions are explored elsewhere \citep{Brewer2013,Portillo2017} and are not the focus of this paper.

\edit1{In this work, we use the term ``maximum-likelihood (ML) photometry'' as an umbrella term for any algorithm that uses (1) a parametric generative model of the data and (2) a noise model to define (3) a likelihood and then (4) estimates fluxes by maximizing this likelihood. The generative model need not be complicated (e.g., the image of a point source may be approximated by a 2-D circular Gaussian) and the noise model can also be simple (e.g., identical errors). In particular, when the noise is assumed to be independent and Gaussian in each pixel maximizing the likelihood amounts to solving a least squares problem (i.e. chi-squared minimization).}

\edit1{ML photometry \textit{does not} include algorithms that use a PSF but do not maximize a likelihood to estimate flux. In a Bayesian statistical framework, for instance, the likelihood can be combined with ``prior beliefs'' to yield a ``posterior'' distribution in flux. Some algorithms return samples from a Bayesian posterior, like The Tractor \citep{Lang2016} in sampling mode or probabilistic cataloging \citep{Brewer2013,Portillo2017} while others use variational inference, like Celeste \citep{Regier201802,Regier201801}.}

This maximum likelihood approach has been implemented in many software packages that are widely used in astronomy, including DAOPHOT/ALLFRAME \citep{Stetson1987,Stetson1994}, DoPHOT \citep{Schechter1993}, HSTphot/DOLPHOT \citep{Dolphin2000}, Photo \citep{Lupton2001}, psphot \citep{2016arXiv161205244M}, The Tractor \citep{Lang2016} in optimization mode, the Hyper Suprime-Cam (HSC) pipeline \citep{2018PASJ...70S...5B}, and the LSST Stack \citep{Juric2017}. While these software packages differ on implementation details (e.g. estimating the sky level, measuring the PSF, and deblending sources), they are similar in how they measure the positions and fluxes of sources. \edit1{They all maximize the likelihood of a PSF (or PSF-convolved model for extended sources) as a function of position and flux} (PSF/model fitting photometry for point/extended sources), possibly alongside aperture photometry at the same positions.

\edit1{Although} \edit1{ML} photometry (in theory) maximizes the SNR (i.e. is most \textit{precise}), that \textbf{does not} mean that it is unbiased (i.e. is most \textit{accurate}) or gives proper errors (i.e. has appropriate \textit{coverage}). Indeed, \edit1{\cite{10.2307/2984505} showed} that maximum-likelihood estimators are generically biased to order ${\rm SNR}^{-2}$. While there are many biases/systematic errors that can appear in photometry (cf. \citealt{Nyland_2017}), in this paper we focus on the biases that arise in \edit1{ML} photometry due to it being a \edit1{maximum-likelihood estimator}. We show that this leads the flux density estimates to generally be biased high, and that it broadly arises from PSF photometry ``over-fitting'' the data \textit{in a way that breaks symmetries}. While this bias is generally small (a $\sim 1\%$ effect at $10\sigma$ for a point source) and likely not of concern for an individual source, its impact may be magnified when using a large population of low signal-to-noise sources.

In this work, we present corrections for the \edit1{flux} bias and consider its implications for \edit1{point sources and extended sources across multiple bands}. We first introduce a simplified version of maximum-likelihood photometry with a single source and Gaussian noise in Section \ref{sec:preliminaries}. We derive the \edit1{flux} bias in the maximum-likelihood flux for a point source and the corresponding \edit1{uncertainty introduced when correcting for it} in Section \ref{sec:bias}. In Section \ref{sec:multi}, we compute the \edit1{flux} bias for multi-band photometry in both the forced photometry and simultaneous fit case. In Section \ref{sec:real}, we show that the \edit1{flux} bias is present in mock data pipelines and SDSS data. \edit1{We discuss our findings in Section \ref{sec:disc}, showing that the flux bias persists when using aperture photometry and that the flux bias is larger for extended sources such as galaxies. We conclude in Section \ref{sec:conc}.}

All the data and code used to create the plots presented in the paper are available online at \url{https://github.com/joshspeagle/phot_bias}. We invite readers to (re-)create their own plots and investigate the nature of this bias for themselves.

\section{Maximum-Likelihood Photometry}
\label{sec:preliminaries}

Consider an $\xdim \times \ydim$ footprint of a pixelated image containing only one point source at some true position $(x^*,y^*)$ with some true flux density $f^*$. Assume the pixel-convolved point spread function (PSF) is constant throughout the footprint and is exactly known. Let the value of the pixel-convolved PSF in pixel $i$ for a point source located at position $(x,y)$ be $p_i(x,y)$. For notational convenience, the values across the whole image will be modeled as the $\xdim\ydim \times 1$ column vector $\psf_{x,y} \equiv \psf(x,y) = \lbrace \dots, p_i(x,y), \dots \rbrace$.

Assume that an estimate of the sky background\footnote{We define the sky background to be any flux \emph{not} coming from the sources that are being cataloged, eg. scattered light from the atmosphere and telescope, dark current, and \edit1{sources below the detection threshold}.} has been subtracted from the footprint and that (residual) sky background in the footprint is Normally distributed with $\xdim\ydim \times 1$ mean column vector $\bflux^*$ and $\xdim\ydim \times \xdim\ydim$ covariance matrix $\cov$, where the noise $\cov$ is known but the mean bias $\bflux^*$ is not. The observed background noise in our pixels $\hat{\bflux} \equiv \lbrace \hat{b}_1, \dots, \hat{b}_{\xdim\ydim} \rbrace$ is then distributed as
\begin{equation}
\hat{\bflux} \sim \mathcal{N}(\bflux^*, \cov)
\end{equation}
where $\mathcal{N}(\mean, \cov)$ is the multivariate Normal distribution with mean vector $\mean$ and covariance matrix $\cov$.

In the case where the background noise is independently and identically distributed (iid) with mean $b^*$ and variance $\sigma^2$ the noise in each pixel follows
\begin{equation}
\hat{b}_i \sim \mathcal{N}(\mu=b^*, \sigma^2=\sigma^2)
\end{equation}
and the mean and covariance become $\bflux^* = \lbrace b^*, \dots, b^* \rbrace$ and $\cov = \diag(\lbrace \sigma^2 \rbrace)$. Although we will often use vector/matrix notation for compactness, we will derive results assuming the iid case throughout the main text. Some corresponding results for the general case are included in Appendix \ref{ap:gen_res}.

Excluding noise from the sky background, the value of the model image for a point source at location $(x,y)$ and with flux density $f$ is
\begin{equation}
\flux(x,y) = f \psf_{x,y}
\end{equation}
The $\xdim\ydim \times 1$ observed flux densities $\hat{\flux}$ within the footprint for our object with true flux density $f^*$ at position $(x^*, y^*)$ are then distributed as
\begin{equation}
\hat{\flux}(x^*, y^*, \bflux^*) = \flux^*(x^*,y^*) + \hat{\bflux} \sim \mathcal{N}(f^*\psf_{x^*,y^*} + \bflux^*, \cov)
\end{equation}
The log-likelihood for a model consisting of a single point source at location $(x,y)$ with flux $f$ and background $\bflux$ is then
\begin{equation}
\ln \mathcal{L}(x,y,f,b) = -\frac{\xdim\ydim}{2} \ln(2\pi\sigma^2) -\frac{1}{2\sigma^2} \sum_i (\hat{f}_i - f p_i(x,y) - b)^2
\end{equation}


When extracting photometry, most often a maximum-likelihood (ML) approach is used. While ML estimators have been widely studied in the statistics literature, we derive some basic results for their application to photometry in this section for completeness. These results are already known in the literature and appear in, for example, \cite{1983PASP...95..163K}.

\subsection{Flux Density}
\label{subsec:ml_flux}

Denote the maximum-likelihood (i.e. best-fit) flux for a given position $(x,y)$ and background $\bflux$ as $f_{\ml}(x,y,\bflux)$. By definition, the partial derivative of the log-likelihood at $f_{\ml}(x,y,\bflux)$ with respect to $f$ is zero such that
\begin{equation}
\partial_f\ln\mathcal{L}(x,y,f_\ml,b) = \frac{1}{\sigma^2} \sum_i (\hat{f}_i - b) p_i(x,y) - f_\ml p_i^2(x,y) = 0
\end{equation}
where $\partial_f \equiv \partial/\partial f$.
This yields
\begin{equation}
\boxed{
f_\ml(x,y,b) = \frac{\sum_i (\hat{f}_i - b) p_i(x,y)}{\sum_i p_i^2(x,y)} \label{eqn:ml_flux}
}
\end{equation}
which is equivalent to using the PSF centered at $(x,y)$ as a matched filter against the background-subtracted image.

We can construct a naive estimate of the error/uncertainty in the maximum-likelihood flux density estimate $\tilde{\sigma}_f(x,y)$ by calculating the (negative inverse of) the second derivative with respect to flux density
\begin{equation}
\tilde{\sigma}_f^2(x,y) \equiv -\left(\partial^2_f\ln\mathcal{L}\right)^{-1} = \frac{\sigma^2}{\sum_i p_i^2(x,y)} \equiv \boxed{\neff(x,y) \times \sigma^2} \label{eqn:flux_err_naive}
\end{equation}
where $\neff(x,y) \equiv 1 / \sum_i p_i^2(x,y)$ is the ``effective area'' of the PSF. This uncertainty is proportional to the background noise $\sigma$ and effective PSF area $\sqrt{\neff(x,y)}$ (i.e. \edit1{the uncertainty} is larger when the PSF is broader). The estimate is naive because it ignores the possible covariances between flux density and other parameters. \edit1{See Appendix \ref{ap:error} for additional details.}

As an example, a circular Gaussian PSF with a standard deviation of $s$ pixels has an effective area of
\begin{equation}
\neff^{\rm G} \rightarrow 4\pi s^2
\end{equation}
in the oversampled limit where $(\xdim, \ydim) \gg s \gg 1$ (i.e. the footprint is large compared to the size of the PSF which is also large compared to the size of a single pixel).

\subsection{Position}
\label{subsec:ml_pos}

We can define the maximum-likelihood positions $(x_\ml,y_\ml)$ the same way by setting the partial derivative of the log-likelihood with respect to $(x,y)$ to 0. Using $x_\ml$ as a representative case we get
\begin{equation}
\partial_x \ln\mathcal{L}(x_\ml,y,f,b) = \frac{f}{\sigma^2} \sum_i \left( (\hat{f}_i - b) - f p_i(x_\ml,y) \right) \partial_x p_i(x_\ml,y) = 0
\end{equation}
Similarly, the naive error/uncertainty in $(x_\ml, y_\ml)$ can be found by calculating the second derivative with respect to position:
\begin{equation}
\tilde{\sigma}_x^2(x_\ml,y,f,b) = \frac{\sigma^2}{f^2} \left( \sum_i \left(\partial_x p_i(x_\ml,y)\right)^2 - \frac{1}{f}\left( (\hat{f}_i - b) - f p_i(x_\ml,y) \right) \partial^2_x p_i(x_\ml,y) \right)^{-1}
\end{equation}

This expression has two components. The first involves the square of the first derivative of the pixel-convolved PSF $(\partial_x p_i)^2$, while the second involves the second derivative of the pixel-convolved PSF $\partial^2_x p_i$ weighted by (fractional) model residuals. Assuming that the residuals are sufficiently small relative to $f$ and the PSF varies sufficiently slowly across the footprint, we can ignore this term and approximate the error as
\begin{equation}
\tilde{\sigma}_x^2(x_\ml,y,f) \approx \frac{1}{f^2} \frac{\sigma^2}{\sum_i \left(\partial_x p_i(x_\ml,y)\right)^2} \equiv \boxed{\seff(x_\ml,y) \times \left(\frac{f^2}{\sigma^2}\right)^{-1}}
\end{equation}
where $\seff(x,y) \equiv 1 / \sum_i \left(\partial_x p_i(x_\ml,y)\right)^2$ is the ``effective smoothness'' of the PSF, analogous to the effective area $\neff(x,y)$. Note that the position error is directly proportional the effective PSF smoothness $\sqrt{\seff(x,y)}$ and inversely proportional to the signal-to-noise ratio (SNR) $f/\sigma$. 

As above, for the case with a circular Gaussian PSF with a standard deviation of $s$ pixels, the effective smoothness is
\begin{equation}
\seff^{\rm G} \rightarrow 8 \pi s^4 = 2 s^2 \neff
\end{equation}
in the oversampled limit for $(\xdim, \ydim) \gg s \gg 1$. This gives a corresponding position error of
\begin{equation}
\label{eqn:position_error}
\sigma_x^2 \rightarrow \frac{2 s^2 \neff^{\rm G} \sigma^2}{f^2} = 2s^2 \left(\frac{f^2}{\tilde{\sigma}_f^2}\right)^{-1}
\end{equation}
where $\tilde{\sigma}^2_f$ is the naive flux density error estimate from \S\ref{subsec:ml_flux}.

\subsection{Background}
\label{subsec:ml_back}

For a given flux density $f$ and position $(x,y)$, we can solve for the maximum-likelihood background $b_\ml$ by again taking the first derivative
\begin{equation}
\partial_b\ln\mathcal{L}(x,y,f,b_\ml) = \frac{1}{\sigma^2}\sum_i \hat{f}_i - f p_i(x,y) - b_\ml = 0
\end{equation}
and setting it to $0$. The ML solution is
\begin{equation}
b_\ml(x,y,f) = \frac{1}{nm} \sum_i \hat{f}_i - f p_i(x,y)
\end{equation}
The associated naive errors then are
\begin{equation}
\boxed{\tilde{\sigma}_b^2 = \frac{\sigma^2}{\xdim\ydim} \equiv \frac{\sigma^2}{A}}
\end{equation}
where $A = \xdim\ydim$ is the area of the footprint.

This result shouldn't be surprising, since it implies that the maximum-likelihood background is just the mean residual between the model $f\psf_{x,y}$ and the data $\hat{\flux}$ in our given $\xdim \times \ydim$ footprint. Since we have assumed a fixed value $b$ across the footprint, this is summed over all the pixels. \edit1{While the background often shows significant spatial variation, approximating it as locally constant in a small area around a source can be useful. For example, psphot fits for a constant local background in its convolved galaxy model fits \citep{2016arXiv161205244M}. To capture spatial variation in larger areas, the background is often approximated by a slowly varying function: for example, hscPipe uses a sixth-order 2D Chebyshev polynomial \citep{2018PASJ...70S...5B}.} Results for the general case where the background is actually a function of a $\bdim$ nuisance parameters $\bparams$ across the footprint are outlined in Appendix \ref{ap:back_gen}. 

\section{Bias in Maximum-Likelihood Estimates}
\label{sec:bias}

\begin{figure}
\includegraphics[scale=0.235]{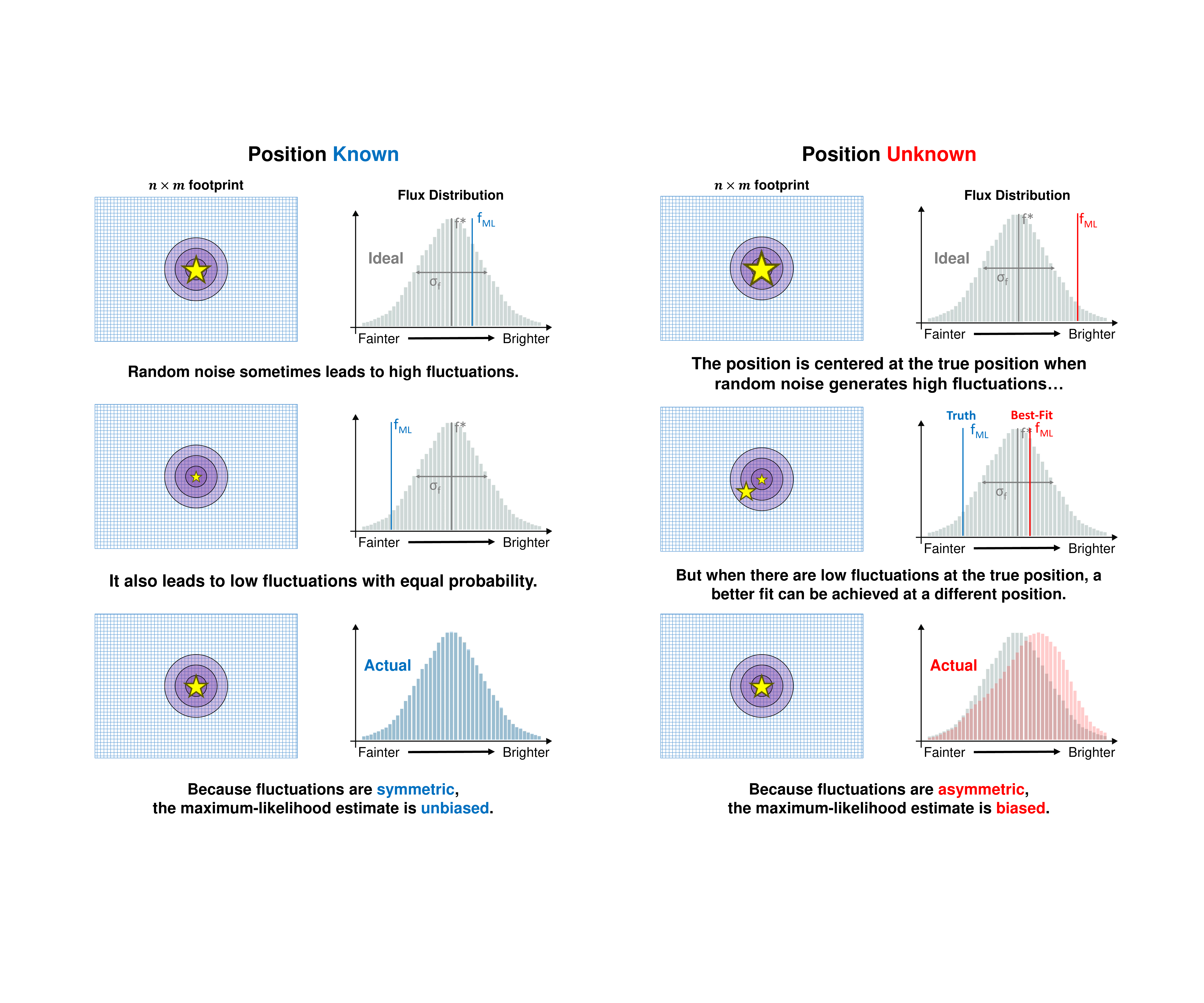}
\vspace{-20mm}
\caption{A schematic illustration of the bias described in \S\ref{sec:bias}. If the true position is known and held fixed, then random noise fluctuations tend to cause the estimated ML flux density $f_\ml$ to randomly fluctuate around the true value $f^*$ with a typical dispersion of $\sigma_f$ (see \S\ref{sec:preliminaries}). Since these fluctuations are \textit{symmetric}, the average estimated flux density is \textit{unbiased}, as shown on the left in blue. Allowing the position $(x,y)$ to vary, however, breaks this symmetry because noise fluctuations tend to draw the ML position $(x_\ml, y_\ml)$ away from the true position $(x^*, y^*)$ to improve the fit. This slightly biases $f_\ml$ against smaller values near the true position and leads to an overall \textit{positive bias} in the estimated flux density, as shown on the right in red. This argument can be generalized to more complex models such as galaxies (see \S\ref{subsec:gal}), which introduce additional ways in which model parameters can ``soak up noise'' in ways that break symmetry.}
\label{fig:schematic}
\end{figure}

While ML estimators are \textit{consistent} estimators (i.e. $f_\ml \rightarrow f^*$ as $f/\sigma_f \rightarrow \infty$) assuming no model mismatch, they are \textit{not} guaranteed to be unbiased at any finite SNR. In this section, we derive estimates of the expected bias between  the ML flux density estimate $f_\ml$ and the true flux density $f^*$ along with its associated variance in the ideal case where $(x^*, y^*, b^*)$ are known (\S\ref{subsec:bias_ideal}) and the general case where it is not (\S\ref{subsec:bias_gen}). A schematic outline of our results is illustrated in Figure \ref{fig:schematic}.

\subsection{Ideal Case}
\label{subsec:bias_ideal}

It is helpful to rewrite the noisy observed flux densities $\hat{\flux}$ in terms of random variables such that
\begin{equation}
\hat{f}_i = f^*p_i(x^*, y^*) + b^* + \sigma Z_i \sim \mathcal{N}(f^*p_i(x^*, y^*) + b^*, \sigma^2) \label{eqn:rv}
\end{equation}
where each $Z_i \sim \mathcal{N}(0, 1)$ is an iid random variable drawn from the standard Normal distribution with mean $\mu=0$ and variance $\sigma^2=1$.
This represents a re-framing of the underlying data generating process: start with our true underlying model $f^*\psf(x^*, y^*) + \bflux^*$ and add on a particular value comprised of $Z_i$ drawn from $\mathcal{N}(0, 1)$ scaled by the standard deviation $\sigma$.

Given equation \eqref{eqn:rv}, the likelihood at the true position $(x^*, y^*)$ and background $b^*$ then reduces to
\begin{equation}
\ln \mathcal{L}(x^*,y^*,f,b^*) = -\frac{A}{2} \ln(2\pi\sigma^2) - \frac{1}{2\sigma^2} \sum_i ((f^*-f)p_i(x^*,y^*) + \sigma Z_i)^2
\end{equation}
Setting the partial derivative $\partial_f \ln \mathcal{L} = 0$ then allows us to write the ML flux density estimate as
\begin{equation}
f_\ml(x^*,y^*,b^*) = f^* + \frac{\sigma}{\sum_i p_i^2(x^*, y^*)} \sum_i Z_i p_i(x^*, y^*) \label{eqn:flux_rv}
\end{equation}

We can rewrite this by noticing that the latter term is actually normally distributed such that
\begin{equation}
\frac{\sigma}{\sum_i p_i^2(x^*, y^*)} \sum_i Z_i p_i(x^*, y^*) \sim \mathcal{N}\left(0, \frac{\sigma^2}{\sum_i p_i^2(x^*, y^*)} = \tilde{\sigma}_f^2(x^*,y^*)\right)
\end{equation}
where the equality comes from equation \eqref{eqn:flux_err_naive}. This implies that the ML flux density estimate is distributed as
\begin{equation}
\boxed{
f_{\ml}(x^*,y^*,b^*) \sim \mathcal{N}(f^*, \sigma_f^2(x^*,y^*))
}
\end{equation}
since the naive error estimate $\tilde{\sigma}_f^2(x^*,y^*)$ is equivalent to the true error $\sigma_f^2(x^*,y^*)$ in the single parameter case \edit1{(see Appendix \ref{ap:error})}. In other words, given the true values of $(x^*, y^*, \bflux^*)$, the ML flux density estimate $f_\ml$ is an \textit{unbiased} estimate of the true flux density $f^*$ with a variance of $\sigma_f^2(x^*, y^*)$ corresponding to the true variance.

\subsection{General Case}
\label{subsec:bias_gen}

Following our random variable notation above, at the true values $(x^*, y^*, f^*, b^*)$ of the position, flux density, and background, respectively, we can rewrite the likelihood of the noisy data as
\begin{equation}
\ln \mathcal{L}(x^*, y^*, f^*, b^*) = -\frac{A}{2} \ln(2\pi\sigma^2) -\frac{1}{2} \sum_i Z_i^2
\end{equation}
where the $Z_i \sim \mathcal{N}(0, 1)$ are again iid normal random variables. The sum of their squares represents the sum of the error-normalized residuals.
We can rewrite this by recognizing that
\begin{equation}
\sum_{i=1}^{\xdim\ydim} Z_i^2 \sim \chi^2_{A}
\end{equation}
which follows a chi-square distribution with $A=\xdim\ydim$ degrees of freedom.

In general, we expect our best-fit parameters to ``absorb'' some of the scatter present in the data since we allow them to vary when we are trying to maximize the likelihood. We can make this more rigorous using \edit1{Cochran's theorem \citep{cochran_1934}}, which implies that the sum of error-normalized residuals for a fit with $p$ free parameters $\params$ will follow
\begin{equation}
(\hat{\flux} - \flux_{\params_\ml})^\T \cov^{-1} (\hat{\flux} - \flux_{\params_\ml}) \sim \chi^2_{A-p}
\end{equation}
with the sum of error-normalized residuals for the ML solution distributed as
\begin{equation}
(\params^*-\params_\ml)^\T \cov_{\params}^{-1} (\params^*-\params_\ml) \sim \chi^2_{p}
\end{equation}
such that their combined sum leaves us with
\begin{equation}
(\hat{\flux} - \flux_{\params_\ml})^\T \cov^{-1} (\hat{\flux} - \flux_{\params_\ml}) + (\params^*-\params_\ml)^\T \cov_{\params}^{-1} (\params^*-\params_\ml) \sim \chi^2_{A}
\end{equation}
since $\chi^2_{i} + \chi^2_{j} \sim \chi^2_{i+j}$.

\subsubsection{Decoupled Background}
\label{subsubsec:back_dec}

There is an asymmetry between the parameters connected to modeling the object and those connected to modeling the background. When modeling an object, $(f, x, y)$ can all modify the model image and provide information on the scale of the PSF. Even as we increase the size of our footprint $(\xdim \times \ydim) \rightarrow (\infty, \infty)$, there is a minimum variance $\sigma_{f,\min}^2$ achievable set by $\neff^{\max}$ and $\sigma$.

In contrast, the background estimate can continually improve as the image becomes larger. This holds true for any finite-parameter background model (see Appendix \ref{ap:back_gen}) as the size of the footprint $A \rightarrow \infty$ becomes infinitely large. This implies that there is a \textit{fundamental difference} between ``object-related'' parameters and ``background-related'' parameters.

In the case where the area of the footprint is substantially larger than the effective area of the PSF (i.e. $A \gg \neff$), we can consider the object parameters effectively decoupled from background parameters. In Appendix \ref{ap:bias}, we show that the freedom in the object position parameters leads to a bias\footnote{Note that this effect is completely independent of (and unrelated to) selection effects such as Malmquist bias that can also cause sources to naturally be biased high close to detection limits/cutoffs. It is also independent of Eddington bias, which causes the number of bright objects to be overestimated when faint objects are more common than bright ones.} in the generalized ML flux density $f_\ml \equiv f_\ml(x_\ml, y_\ml)$ relative to the ideal (unbiased) ML flux density $f_\ml^* \equiv f_\ml(x^*, y^*)$ in \S\ref{subsec:bias_ideal} that to leading order goes as
\begin{equation}
f_\ml^* \approx f_\ml \left( 1 - \frac{X_{2}^2}{2}\frac{\tilde{\sigma}^2_{f_\ml}}{f_\ml^2} \right) \label{eqn:bias_ideal}
\end{equation}
where $X^2_2 \sim \chi^2_2$ is a random variable drawn from the chi-square distribution with 2 degrees of freedom (which is determined by the 2 parameters $(x, y)$ used to fit for the position). This leads to a fractional bias of
\begin{equation}
\boxed{
\frac{\delta_{f_\ml}}{f_\ml} \equiv 1 - \frac{\mathbb{E}[f_\ml^*]}{f_\ml} \approx \frac{\tilde{\sigma}^2_{f_\ml}}{f_\ml^2}
}
\label{eqn:bias_point_source}
\end{equation}
where $\mathbb{E}[f_\ml^*]$ is the expectation value of $f_\ml^*$. This shows that $f_\ml$ is biased \textit{high} relative to the true underlying flux density $f^*$. See Appendix \ref{ap:bias} for a more detailed derivation involving higher-order terms and \ref{ap:bias_tensor} for an alternate derivation using bias tensors. This bias is the same as the ``gradient bias'' identified by \cite{2007MNRAS.380..199I} and the ``noise bias'' identified by \cite{2012MNRAS.425.1951R}. A related derivation was done by \cite{2010A&A...523A...7G}, which neglected the correlation between noise and the maximum likelihood position and obtained a different result.  The authors of that paper confirm that when they include the noise-position correlation, they agree with our result (private communication). \edit1{Note that unlike the ML flux density, the \textit{mean} flux density $f_{\rm mean}$ (derived from, e.g., Markov Chain Monte Carlo methods) is in fact an \textit{unbiased} estimator of the true flux. See Appendix \ref{ap:flux_mean} for additional details.}

This result gives a straightforward procedure to approximately ``de-bias'' $f_\ml$ using equation \eqref{eqn:bias_ideal}. Doing so, however, increases the variance in the measurement to first-order by
\begin{equation}
\frac{\mathbb{V}[f_\ml^*]}{\tilde{\sigma}_{f_\ml}^2} \approx \frac{\sigma^2_{f_\ml}}{f_\ml^2}
\end{equation}
since the exact bias for any individual measurement is not known. This is an example of the \textbf{bias-variance trade-off} and leads to an increase in the total effective error following
\begin{equation}
\tilde{\sigma}_{f_\ml^*} = \sqrt{\tilde{\sigma}_{f_\ml}^2 + \mathbb{V}[f_\ml^*]} \approx \tilde{\sigma}_{f_\ml} \left(1 + \frac{1}{2}\frac{\mathbb{V}[f_\ml^*]}{\tilde{\sigma}_{f_\ml}^2}\right) = \boxed{\tilde{\sigma}_{f_\ml} \left(1 + \frac{1}{2}\frac{\tilde{\sigma}^2_{f_\ml}}{f_\ml^2}\right)}
\end{equation}

The magnitude of this bias for PSF photometry is generally small but not totally negligible: a nominally $10\sigma$ source (i.e. $f_\ml = 10\tilde{\sigma}_{f_\ml}$) incurs a 1\% bias and 0.5\% error underestimate. We will return to this in \S\ref{subsec:gal} when we examine the behavior of $f_\ml$ when modeling extended objects.

\begin{figure}
\includegraphics[scale=0.3]{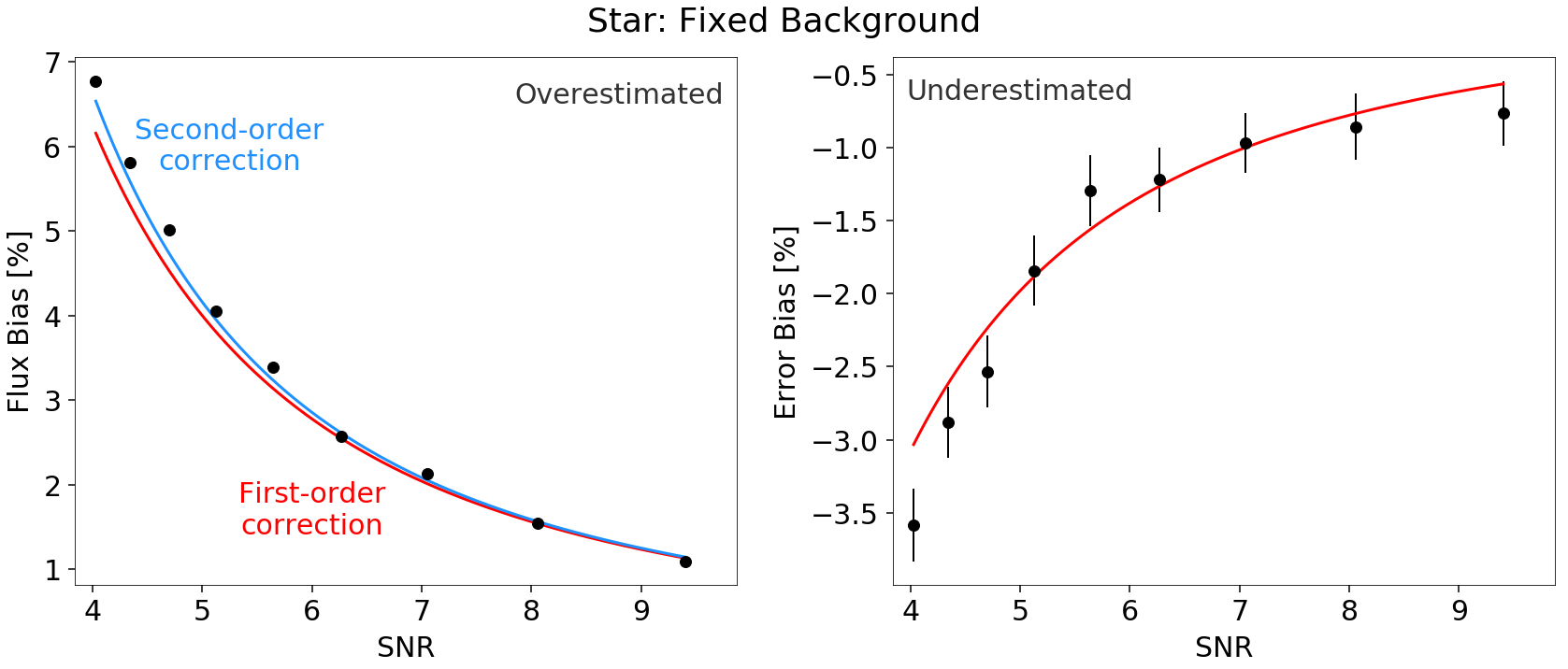}
\caption{\textbf{Left:} The bias in the estimated maximum-likelihood flux density for a point source assuming a fixed background relative to the true value. First (red) and second-order (blue) analytic predictions (Appendix \ref{ap:bias}) are compared with results from simulated images. The flux estimate is biased high because the position parameters will move to better fit the noise in the data, increasing the log-likelihood and estimated flux. To leading order in SNR, the fractional flux bias is ${\rm SNR}^{-2}$. \textbf{Right:} The bias in the derived errors after correcting for the mean bias in the ML flux. Since the exact bias for any individual measurement is not known, subtracting the average bias increases the error by $1/2 \times {\rm SNR}^{-2}$. This is an example of the \textit{bias-variance tradeoff} where decreasing the bias increases the variance.}
\label{fig:star_fixed_back}
\end{figure}

We test our analytic predictions by creating a set of simulated point-source images and running maximum-likelihood photometry on them. Our simulated images have a point-source with a circular Gaussian PSF of $\sigma=2\;\rm{pixels}$ in the center of a $101\times101$ pixel image with iid Gaussian noise in each pixel. We simulate sources of nine fluxes ranging from $4.0\sigma$ to $9.4\sigma$, evenly spaced in $1/SNR$. For each flux, we create 100,000 different simulated images. We solve for the ML parameters using the Trust Region Reflective algorithm (as implemented in scipy), initialized with the true parameters to ensure that the optimizer finds the global maximum. Figure \ref{fig:star_fixed_back} shows that the mean flux bias and flux errors from these simulated images agree well with our predictions.

\begin{figure}
\includegraphics[scale=0.16]{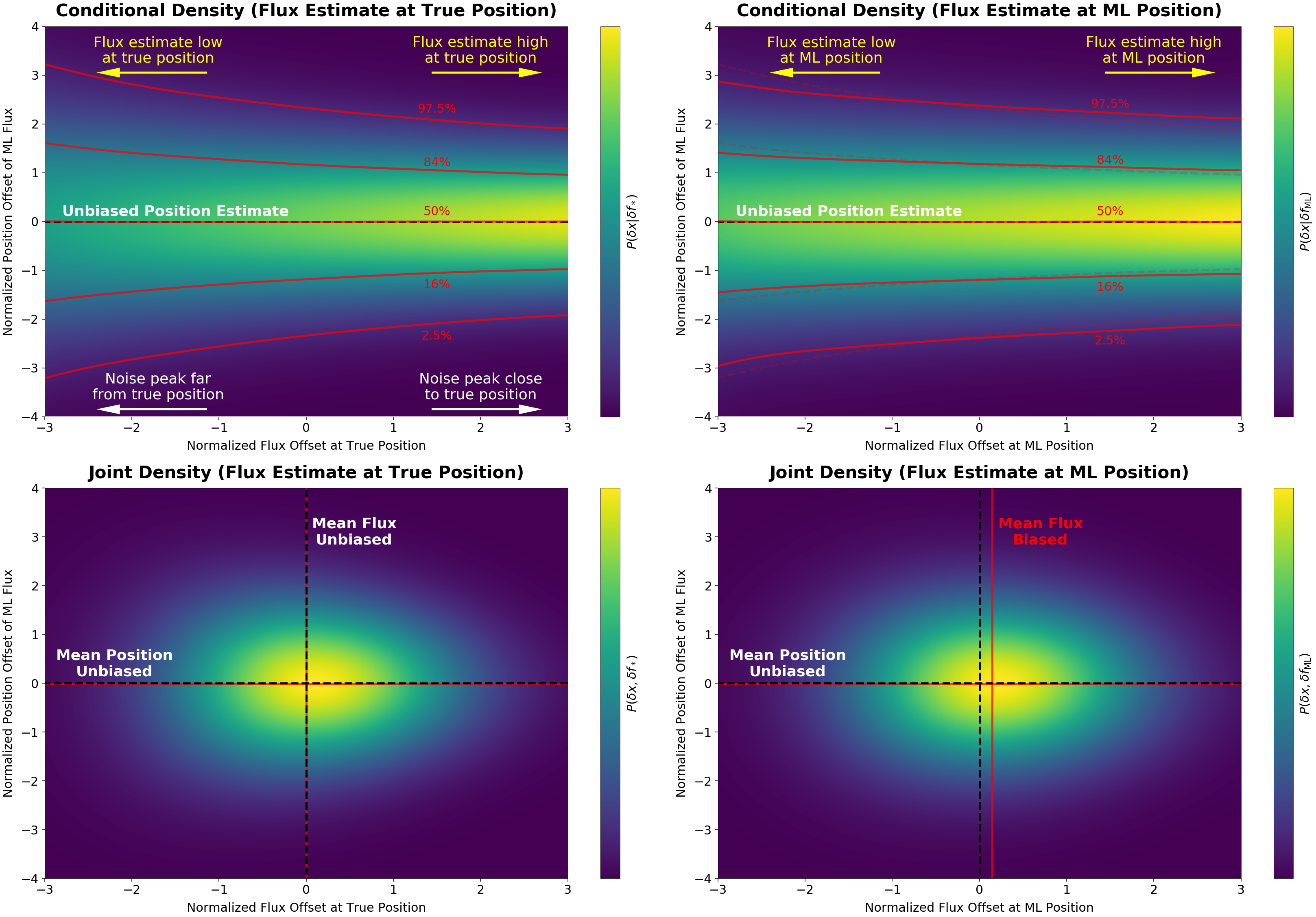}
\caption{An illustration of the ``centering bias'' described in Figure \ref{fig:schematic}. When random fluctuations lead to ``noise peaks'' near the true position, the maximum-likelihood position $(x_\ml, y_\ml)$ will remain close to the true position $(x^*, y^*)$ but the flux density $f_\ml$ will be overestimated. When random fluctuations generate noise peaks further from the true position, however, the fit follows them away from the true position (top panels). This process systematically biases the distribution of $f_\ml(x_\ml, y_\ml)$ at the maximum-likelihood position to larger values compared to the distribution of $f_\ml(x^*, y^*)$ values that \textit{would be} estimated at the true position (bottom panels).The general behavior in Figure \ref{fig:star_fixed_back} describing the mean bias arises from averaging over these two behaviors.}
\label{fig:star_varpos}
\end{figure}

\begin{figure}
\includegraphics[scale=0.25]{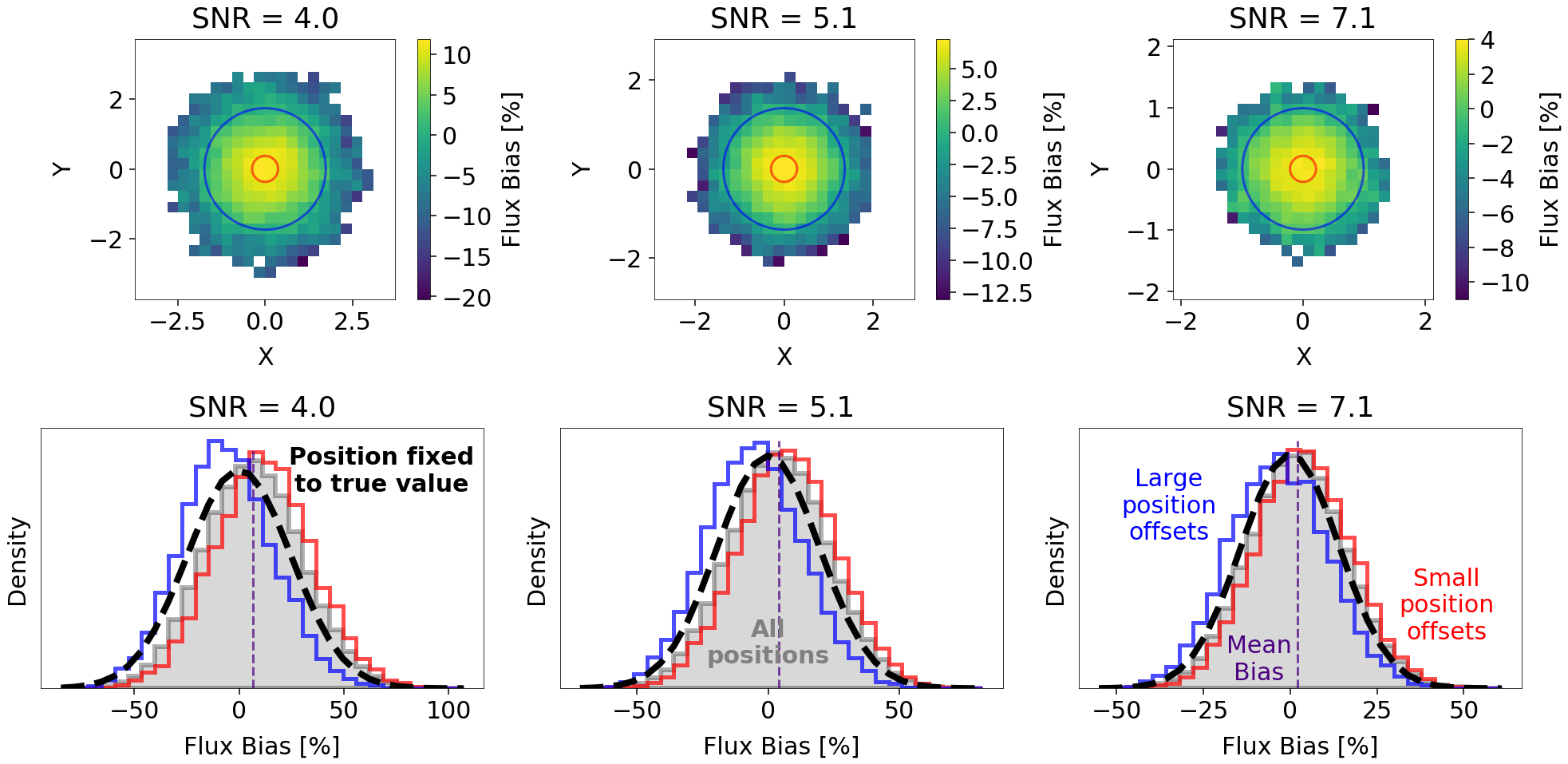}
\caption{The impact of ``centering bias'' on the estimated maximum-likelihood flux density $f_\ml$ as a function of signal-to-noise ratio (SNR). \textbf{Top:} The bias in $f_\ml$ for a point source, assuming a fixed background, relative to the true value as a function of position $(x, y)$ at various signal-to-noise ratios (SNR). In all cases, $f_\ml$ is biased high near the true position and low at the outskirts with amplitudes based on the SNR.
\textbf{Bottom:} The corresponding distribution of $f_\ml$ across all positions (gray) compared with those with small position offsets (red, extracted inside the red circles in the top panels) and large position offsets (blue, extracted outside the blue circles in the top panels). The (unbiased) distribution that would arise if the position $(x,y)$ was fixed to the true value $(x^*, y^*)$ is shown as the thick black dashed curve. Because the position is left free, the fit is allowed to move away from the true position whenever noise fluctuations would tend to lead to smaller inferred $f_\ml$ at a given position. This systematically ``removes'' low $f_\ml$ estimates derived near the true position, which are subsequently biased high. At larger position offsets, the model has less overlap with the true PSF, which in general biases $f_\ml$ low. The general behavior in Figure \ref{fig:star_fixed_back} describing the mean bias (shown in purple) arises from averaging over these two behaviors.}
\label{fig:star_fixed_back_pos}
\end{figure}

It is crucial to note this bias \textit{does not} by itself arise from the fact that ML estimators tend to ``overfit'' the data when they use more parameters than are justified by the data. If that were the case, we should not have found that the ML estimate at the true position $f_\ml$ was actually unbiased because the fitted parameters are exactly those used in the data-generating process. Instead, this bias arises because of the \textit{way} in which this overfitting occurs. At the true position, $f_\ml$ is allowed to adjust to the noise, but it does so in a symmetrical way: the noise fluctuations in the image are symmetric, and so $f_\ml$ is just as likely to fluctuate upwards relative to $f^*$ as it is to fluctuate downwards. Once the position $(x, y)$ is allowed to vary, however, the model can move the source to absorb nearby positive noise. This breaks the symmetry from earlier: the source will tend to stay in the correct position with an overestimated $f_\ml$ when the noise fluctuates upwards around the true position $(x^*, y^*)$, but will try to move position and absorb nearby positive noise when the noise fluctuates downwards around $(x^*, y^*)$. This position-dependent behavior of $f_\ml$ is broadly illustrated in Figure \ref{fig:schematic} and shown in more detail in Figures \ref{fig:star_varpos} and \ref{fig:star_fixed_back_pos}. Averaging over this behavior as a function of position then gives the results derived above. We note that for sources that can be confidently detected (ie. ${\rm SNR} > 5$), the ML positions only move from the true positions by fractions of a pixel (see Equation \eqref{eqn:position_error}). Thus the bias arises because the noise modifies the global maximum in the likelihood function, not because it creates a new local peak.

\subsubsection{Coupled Background}
\label{subsubsec:back_coup}

The result we derived above holds if the image is sufficiently large that background estimation is effectively decoupled from modeling the actual object. In the case where this is not true (i.e. $A \not\gg \neff$), we instead need to consider how background estimation is covariant with our model parameters. \edit1{For example, in extragalactic deep fields, $\neff$ is larger than it would be for point sources because it is the effective area of a galaxy profile convolved with the PSF, while the footprint area $A$ is limited by spatial variations in the PSF.}
The mixed 2nd-order derivatives of the log-likelihood give
\begin{align}
\partial_f \partial_b \ln \mathcal{L}(x,y) &= -\frac{1}{\sigma^2} \sum_i p_i(x,y) = -\frac{1}{\sigma^2} \\
\partial_x \partial_b \ln \mathcal{L}(x,y) &= -\frac{f}{\sigma^2} \sum_i \partial_x p_i(x,y) \approx 0
\end{align}
assuming that $p_i(x,y)$ is oversampled and roughly symmetric in $x$. As shown in \edit1{Appendix \ref{ap:error}}, the contribution from the mixed partial with respect to $f$ and $b$ is expected to contribute to a fractional underestimate of the variance proportional to the ratio $A/\neff$ of the area of the footprint versus the effective area of the PSF. This gives a modified variance of
\begin{equation}
\sigma_f^2(x,y) = \frac{\neff(x,y) \times \sigma^2}{1 - \frac{\neff(x,y)}{A}} = \frac{A}{A-\neff(x,y)} \times \tilde{\sigma}_f^2
\label{eqn:flux_error_back_cov}
\end{equation}
since by construction $A \geq \neff$. Note that we have dropped $\tilde{\sigma}_f$ since $\sigma_f$ is now the ``true'' uncertainty of our ML estimator that takes into account the relative coupling between the ML background estimate $b_\ml$ and the parameters used to model our object $(f_\ml, x_\ml, y_\ml)$.

Substituting in $\sigma_f$ for $\tilde{\sigma}_f$ into our expressions from \S\ref{subsubsec:back_dec} then modifies the effective SNR to give
\begin{equation}
\boxed{
\frac{\delta_{f_\ml}}{f_\ml} \approx \frac{{\sigma}^2_{f_\ml}}{f_\ml^2} \,,~ \frac{\mathbb{V}[f_\ml^*]}{\sigma_{f_\ml}^2} \approx \frac{\sigma^2_{f_\ml}}{f_\ml^2} 
}
\end{equation}
When the effective size of the footprint is large compared to the PSF, then $A / (A - \neff) \rightarrow 1$ and the background is effectively decoupled from the modeling of the object, leading to our results in \S\ref{subsubsec:back_dec}. When the effective size becomes more comparable to the PSF, then $A / (A - \neff) \rightarrow \infty$ and the covariance between the background and flux density dominate the error budget. This has the effect of increasing the expected bias by decreasing the effective SNR.

While a coupled background worsens the bias that arises when the source position is unknown, no bias arises if the true source position is known and the background is uniform. In this case, the model image is linear in all the free parameters ($f$ and $b$) and so no bias is incurred. The covariance between the background level and flux density of the source will still inflate the uncertainty in the flux density, as shown in equation \eqref{eqn:flux_error_back_cov}.

\begin{figure}
\includegraphics[scale=0.3]{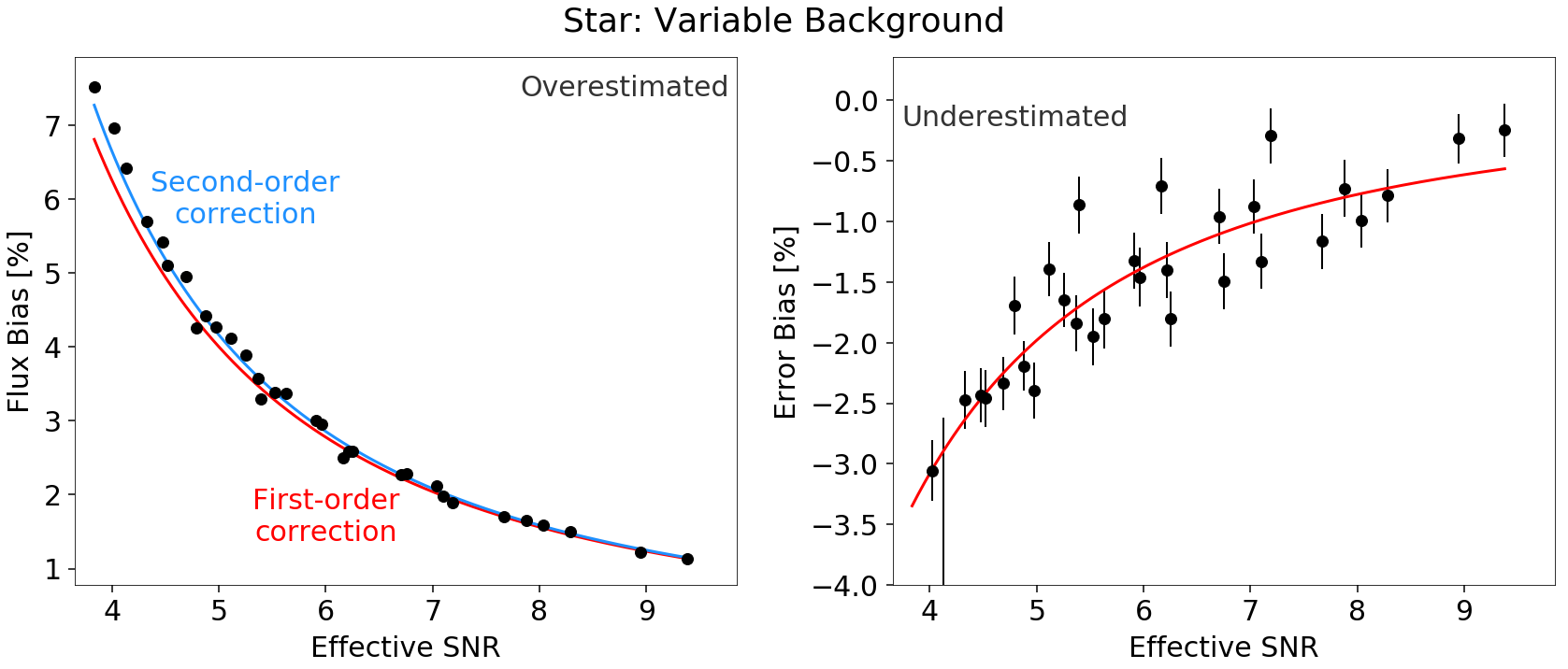}
\caption{As Figure \ref{fig:star_fixed_back}, but now with a variable background and computed over various image sizes. The effective signal-to-noise ratios (SNRs) decrease due to the covariance between the background and the estimated flux density as described in \S\ref{subsubsec:back_coup}, which depend on the relative size of the image $A$ compared to the effective size of the point-spread function $\neff$. The analytic predictions computed using these lower effective SNRs still model the data well.}
\label{fig:star_var_back}
\end{figure}

We augment the set of simulation images used to generate Figure \ref{fig:star_fixed_back} by considering five different image sizes approximately evenly distributed in $1/A$: 11, 13, 15, 23, and 101 pixels. The effective SNR of a given flux will decrease as image size decreases. In Figure \ref{fig:star_var_back}, we show that our analytic predictions in terms of the effective SNR hold in these simulated images.

We want to note that while this effect is conceptually useful going forward, in practice the impact is extremely small. For instance, in SDSS the background is determined in patches of $256\times256$ pixels.\footnote{\url{https://www.sdss.org/dr14/algorithms/sky/}} With a median seeing of $1.32\arcsec$ in $r$ band\footnote{\url{https://www.sdss.org/dr14/imaging/other_info/}} and pixel size of $0.4\arcsec$ \citep{1538-3881-116-6-3040}, $\neff/A = 4.4\times10^{-4}$. Similarly, in LSST the background will be determined in patches of $256\times256$ or $512\times512$ pixels.\footnote{\url{https://confluence.lsstcorp.org/display/LSWUG/Measurement+in+the+LSST+Stack}} With a median seeing of $0.7\arcsec$ in $r$ band and pixel size of $0.2\arcsec$ \citep{2009arXiv0912.0201L}, $\neff/A = 4.3\times10^{-4}$ if the background is determined using $256\times256$ pixel patches. A significantly larger effect is present in unWISE \citep{Schlafly_2019}, which estimates the sky background in much smaller ($20\times20$) pixel regions: with a PSF FWHM of $6\arcsec$ and a pixel scale of $2.75\arcsec$ \citep{2010AJ....140.1868W}, $\neff/A = 2.8\%$.

\section{Extension to Multi-Band Fitting}
\label{sec:multi}

We now examine the case where an object is modeled in multiple bands. We will consider two cases. The first (\S\ref{subsec:phot_single}) is where the object is ``detected'' in a single band, after which the position is fixed across all the bands. This is analogous to surveys such as \edit1{the Hyper Suprime-Cam Subaru Strategic Program Survey (HSC-SSP; \citealt{2018PASJ...70S...4A}), which prefers $i$-band as the ``reference band'' for multi-band object detections \citep{2018PASJ...70S...5B}.} The second (\S\ref{subsec:phot_unforced}) is where the object is modeled simultaneously across all bands. This is somewhat equivalent to \edit1{detecting an object in a stacked (PSF-matched) image constructed from all the bands as proposed by \cite{Szalay_1999}, as is done by the Wide-field Infrared Survey Explorer \citep{2010AJ....140.1868W}, the Cosmological Evolution Survey \citep{2016ApJS..224...24L}, and the Dark Energy Survey \citep{2018PASP..130g4501M}.}

\subsection{Single-band Detection}
\label{subsec:phot_single}

Let's assume that our object is detected in band $D$, after which the ML position $(x_\ml, y_\ml)=(x_D^\ml, y_D^\ml) \equiv (x_D, y_D)$ is fixed. As shown in \S\ref{sec:bias}, we expect that the ML flux density estimate in this band to be overestimated by an amount based on the PSF-normalized SNR. Following our previous assumption that the likelihood is multivariate normal around the ML parameters $\params_\ml$, we would expect the values for our ML flux density estimates $f_{j,\ml}$ in other band $j \neq D$ to be
\begin{equation}
f_{j,\ml}(x_D,y_D,b_j^*) \approx \frac{\sum_i (f_k^* p_i(x^*, y^*) + \sigma_j Z_i) p_i(x_D, y_D)}{\sum_i p_i^2(x_D, y_D)}
\end{equation}
where $Z_i$ is again an iid normally distributed random variable, $\sigma_j$ is the noise in band $j$, and we have again assumed that background $b_j$ is known and fixed to the true value $b_j^*$. This result is analogous to equation \eqref{eqn:flux_rv}, except that we have assumed a mismatch in the position between our model at $(x_D, y_D)$ and the source at $(x^*, y^*)$.

The expectation value of $f_{j,\ml}$ assuming $(x_D, y_D)$ is fixed is
\begin{equation}
\mathbb{E}[f_{j,\ml} \,|\, x_D, y_D] \approx f_j^* \times \frac{\sum_i p_i(x^*, y^*) \, p_i(x_D, y_D)}{\sum_i p_i^2(x_D, y_D)}
\end{equation}
Since we expect the mismatched PSF term to be smaller than the matched PSF term $\sum_i p_i(x^*, y^*) p_i(x_D, y_D) \leq \sum_i p_i^2(x_D, y_D)$, this implies that $\mathbb{E}[f_{j,\ml} \,|\, x_D, y_D] \leq f_j^*$ so that our ML flux densities are underestimated. Note that this bias tends to zero as $(x_\ml, y_\ml) \rightarrow (x^*, y^*)$, again confirming that our ML estimator is consistent in the limit of infinite SNR.

Subsequently taking the expectation value over position then gives the general expression
\begin{equation}
\mathbb{E}[f_{j,\ml}] \approx f_j^* \times \iint \frac{\sum_i p_i(x^*, y^*) \, p_i(x_D, y_D)}{\sum_i p_i^2(x_D, y_D)} \,P(x_D, y_D \,|\, \params^*, \cov_{\params}) \,dx_D\,dy_D
\end{equation}
where $P(x_D, y_D \,|\, \params^*, \cov_{\params})$ is a 2-D multivariate normal distribution for $(x, y)$ (see Appendix \ref{ap:error}). While this integral does not have an analytic solution, since we expect the bias to increase as $(x_D, y_D)$ becomes progressively more offset from $(x^*, y^*)$, performing an average over possible ML positions further away from the true position should not be able to change our overall bias from an underestimate to an overestimate. Therefore, we arrive at the general conclusion that
\begin{equation}
\mathbb{E}[f_{j,\ml}] \lesssim f_j^*
\end{equation}

In other words, while our ML flux density estimates tend to be \textit{overestimated} in the detection band, they will tend to be \textit{underestimated} in all other bands. The severity of this (reverse) bias depends on the exact properties of the PSF in each band relative to the detection band (which establishes the ML position and associated covariances).

\begin{figure}
\includegraphics[scale=0.2]{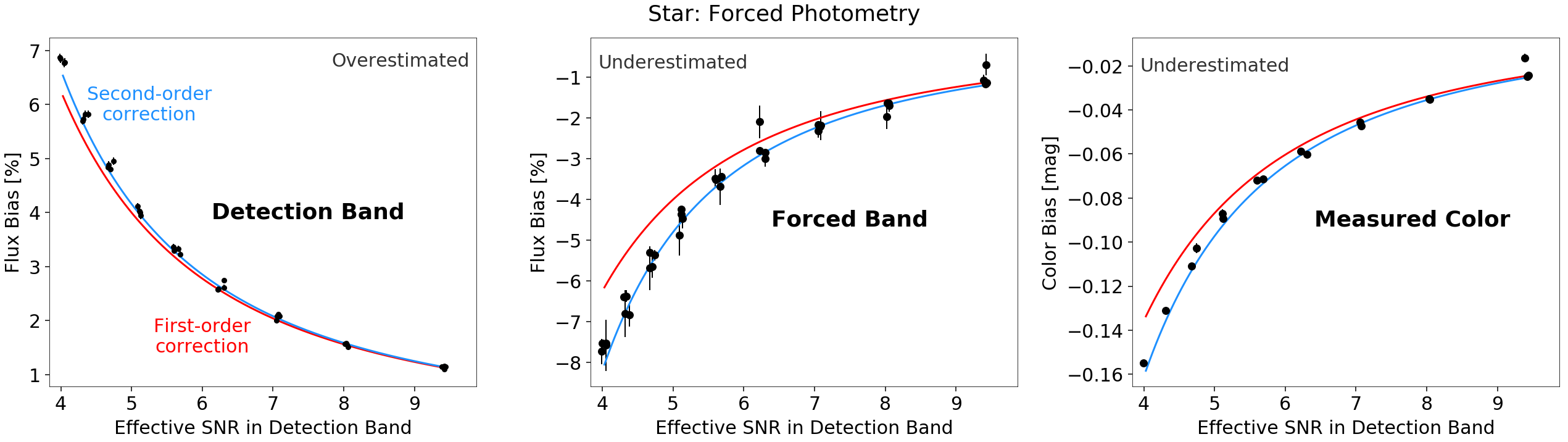}
\caption{The bias in the estimated maximum-likelihood flux density for a point source assuming a fixed background relative to the true value in the \textit{detection} band $f_D$ (left) and the forced band $f_{j,\ml}$ (middle) along with the associated color $C=-2.5\log(f_{D}/f_{j,\ml})$ (right). The first and second-order analytic predictions from Appendix \ref{ap:bias} and \S\ref{subsec:phot_single} are shown in red and blue, respectively. The estimate in the detection band $f_D$ is biased high because the position parameters will move to better fit the noise in the data (see Figure \ref{fig:star_fixed_back}). The estimate in the forced band $f_{j,\ml}$ is biased low since the maximum-likelihood position in the detection band $(x_D, y_D)$ is offset from the true position. The biases in the detection band and forced band compound when considering the color between the two bands. To leading order in the signal-to-noise ratio (SNR) in the \textit{detection} band, the fractional flux bias in the forced band is $-{\rm SNR}_D^{-2}$ and the bias in color is $-5/\ln 10 \times {\rm SNR}_D^{-2}$.}
\label{fig:forced_photometry_bias}
\end{figure}

In the particular case where we have a circular Gaussian PSF with a standard deviation of $s$ pixels, we can evaluate these biases explicitly. At a fixed offset $r_D^2 = (x_D - x^*)^2 + (y_D - y^*)^2$, the expected value of $f_{j,\ml}$ is
\begin{equation}
\mathbb{E}[f_{j,\ml} \,|\, r_D^2] = f_j^* \times \exp\left(\frac{-r_D^2}{4 s^2}\right)
\end{equation}
Assuming that the covariance between $x_D$ and $y_D$ is small such that we can approximate the errors as $\sigma_{D,x}^2 = \sigma_{D,y}^2$, the error-normalized offset will be distributed as
\begin{equation}
\frac{(x_D - x^*)^2}{\sigma_{D,x}^2} + \frac{(y_D - y^*)^2}{\sigma_{D,y}^2} = \frac{r_D^2}{\sigma_{D,x}^2} \sim \chi^2_2
\edit1{\quad\Rightarrow\quad r_D \sim \sigma_{D,x} \times \chi_2}
\end{equation}
This implies that
\begin{equation}
r_D \sim {\rm Rayleigh}(\sigma_{D,x})
\end{equation}
where ${\rm Rayleigh}(\sigma_{D,x})$ is the Rayleigh distribution with scale parameter $\sigma_{D,x}$. \edit1{This is simply the distribution of the magnitude of a 2D vector whose components are drawn from Gaussians centered at zero and arises because we assume the position errors are Gaussian}. Marginalizing over $r_D$ then yields
\begin{equation}
\label{eqn:detect_underest_gaussian}
\mathbb{E}[f_{j,\ml}] = f_j^* \times \frac{2 s^2}{2 s^2 + \sigma_{D,x}^2} 
\end{equation}
Assuming a Gaussian PSF in the detection band with standard deviation $s_D$ and $\sigma_{x,D}^2 = 8 \pi s_D^4 \sigma_D^2 / f_D^2 \approx 2 s_D^2 \sigma_{f,D}^2 / f_D^2$, we get
\begin{equation}
\mathbb{E}[f_{j,\ml}] \approx f_j^* \left(1+\frac{s_D^2}{s^2} \frac{\sigma_{f,D}^2}{f_D^2}\right)^{-1}
\end{equation}
This corresponds to a fractional bias of
\begin{equation}
\boxed{
\frac{\delta_{f_{j,\ml}}}{f_{j,\ml}} \equiv 1 - \frac{f_j^*}{\mathbb{E}[f_{j,\ml}]} \approx -\frac{s_D^2}{s^2} \frac{\sigma_{f,D}^2}{f_D^2}
}
\label{eqn:bias_forced_phot}
\end{equation}

In practice, we find that to properly model the bias at lower SNR requires incorporating a slightly higher-order Taylor expansion of our results (i.e. going from 2nd to 4th-order). As above, this expansion in general is non-trivial. However, it can be evaluated explicitly for circular Gaussian PSFs, as shown in Appendix \ref{ap:flux_taylor}. Including this additional term then gives
\begin{equation}
\frac{f_j^*}{\mathbb{E}[f_{j,\ml}]} - 1 \approx - \frac{s_D^2}{s^2} \frac{\tilde{\sigma}_{f_D}^2}{f_D^{*2}} - \left[\frac{7 s_D^2}{s^2} - \frac{s_D^4}{s^4}\right] \frac{\tilde{\sigma}_{f_D}^4}{f_D^{*4}}
\end{equation}

We test our predictions for forced photometry by creating a set of simulated point-source images in two bands, running maximum-likelihood photometry on one band (the detection band) and forced photometry on the other (the forced band). Our simulated images have a point-source with a circular Gaussian PSF of $\sigma=2\;\rm{pixels}$ in the center of a $101\times101$ pixel image in each band with iid Gaussian noise in each pixel. We simulate sources with nine fluxes ranging from $4.0\sigma$ to $9.4\sigma$ in the detection band, evenly spaced in $1/SNR$. For each detection band flux, we consider four different forced band fluxes: $1\times$, $2\times$, $4\times$, and $8\times$ fainter than the flux in the detection band. For each flux combination, we create 100,000 different simulated images. Figure \ref{fig:forced_photometry_bias} shows that the flux in the detection band is overestimated and the flux in the forced band is underestimated, both by a fraction depending on the SNR in the detection band, as well as the bias on the measured color $C=-2.5\log(f_{D}/f_{j,\ml})$.

\subsection{Unforced Photometry in All Bands}
\label{subsec:phot_unforced}

\begin{figure}
\includegraphics[scale=0.2]{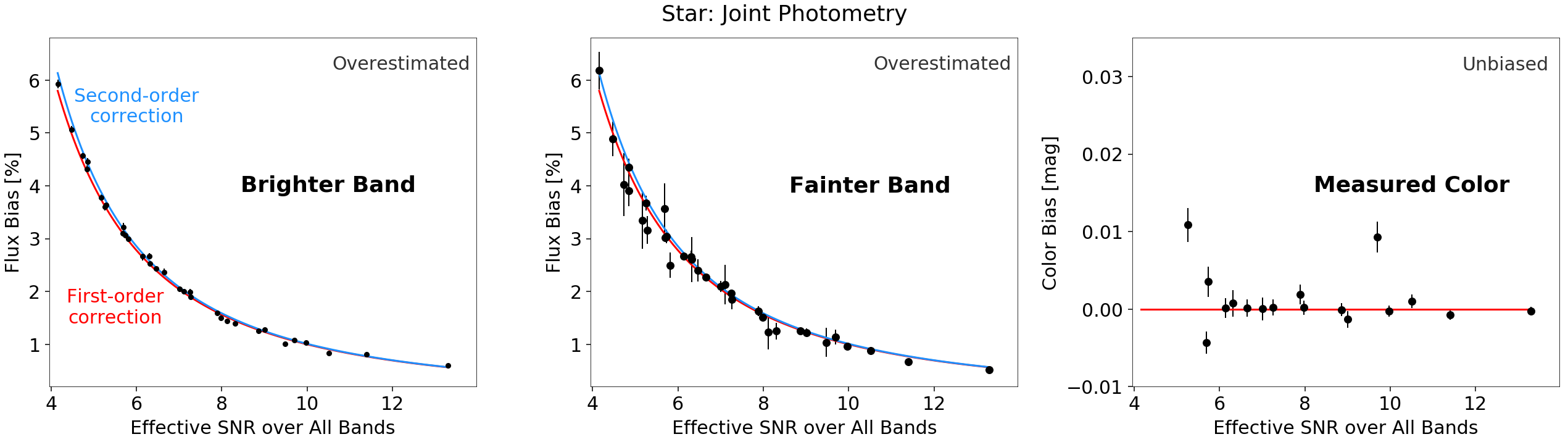}
\caption{As Figure \ref{fig:forced_photometry_bias}, but for simultaneous two-band photometry with a common $(x,y)$ position. The position parameters move to better fit the noise in both bands, splitting the increase in the log-likelihood and estimated flux between both bands. If the two bands have the same PSF size, the fractional flux bias in both bands is determined by the combined SNR, regardless of color. Thus, the measured color is not affected by these flux biases, unlike the forced photometry case shown in Figure \ref{fig:forced_photometry_bias}. To leading order in the combined signal-to-noise ratio (SNR), the fractional flux bias in both bands is ${\rm SNR}_{\rm tot}^{-2}$.}
\label{fig:simultaneous_fit_bias}
\end{figure}

When the object is modeled simultaneously across bands (i.e. with a common $(x,y)$ position), using Cochran's theorem reveals that the sum of fluxes across bands will be biased high but does not show how the flux in individual bands is biased. To calculate the bias in each band, we use the bias tensor formulation introduced by \cite{10.2307/2984505}, which shows that the leading-order term in this bias for parameter $s$ is
\begin{equation}
\bias_s(\params_\ml) = \sum_{r,t,u} (\fisher^{-1}(\params_\ml))_{rs} \, (\fisher^{-1}(\params_\ml))_{tu} \, (\btensor(\params_\ml))_{rtu}
\end{equation}
where 
\begin{equation}
(\btensor(\params_\ml))_{rtu} \equiv \mathbb{E}_{\mathbf{D}}\left[\frac{1}{2} \partial_r\partial_t\partial_u\ln \mathcal{L}(\params_\ml) + (\partial_t \ln \mathcal{L}(\params_\ml)) (\partial_r\partial_u \ln \mathcal{L}(\params_\ml)) \,\middle|\, \params_\ml \right]
\end{equation}
is the \textbf{bias tensor} and $\mathbb{E}_\mathbf{D}[\cdot|\params_\ml]$ is the expectation value with respect to the data $\mathbf{D}$ for $\params_\ml$ fixed.


\begin{figure}
\includegraphics[scale=0.19]{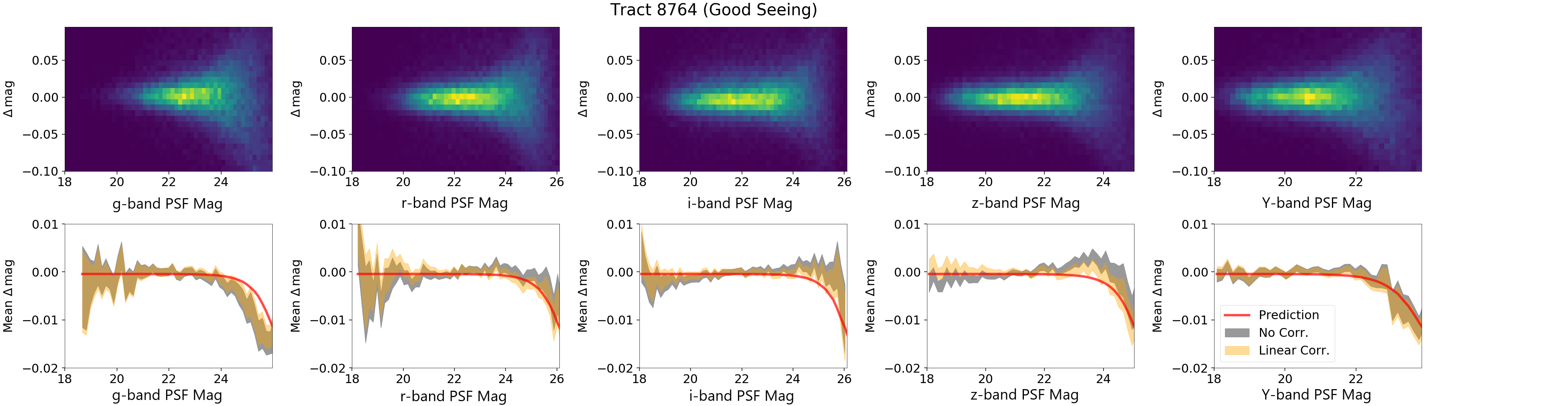}
\caption{\textbf{Top}: The distribution of magnitude offsets in the $grizy$ bands from HSC-SSP SynPipe artificial star tests as a function of input (true) magnitude for a tract with good seeing ($< 1 \arcsec$). \textbf{Bottom:} The mean magnitude offsets as a function of magnitude from the top panel with (yellow) and without (gray) a simple linear correction to account for observed systematic trends. The mean predicted first-order analytic correction (${\rm SNR}^{-2}$) is shown in red. The mean magnitude offsets have been shifted to accommodate zero-point differences. Note the difference in scale between the top and bottom panels, highlighted the subtlety of the derived bias. Our analytic prediction of the bias tracks the data well, especially once the observed linear bias has been accounted for.}
\label{fig:hsc_good}
\end{figure}

The derivation of the bias in $f_\ml$ with respect to $f_\ml^*$ using bias tensors in the single-band case is outlined in Appendix \ref{ap:bias_tensor}. There, we show that
\begin{equation}
\bias_f(\params_\ml) = \tilde{\sigma}_{f_\ml}^2 \sum_{i\in \{x,y\}} \sigma_x^2 \, (\btensor(\params_\ml))_{fii}
\label{eqn:biastensor_backfixed1}
\end{equation}
where
\begin{equation}
\label{eqn:biastensor_elements1}
(\btensor(\params_\ml))_{fxx} = (\btensor(\params_\ml))_{fyy} = \frac{1}{2 \sigma_x^2 f}
\end{equation}
When using multiple bands, the bias for band $i$ has the same terms as equation \eqref{eqn:biastensor_backfixed1} for the single band case, but $\sigma_x^2=(\fisher^{-1})_{xx}$ is smaller because all bands help constrain the position. However, in the bias tensor (equation \ref{eqn:biastensor_elements1}), the $\sigma_x^2$ is the uncertainty in position that would have been obtained using only band $i$. If all bands have Gaussian PSFs with widths $s_j$, then the flux bias for band $i$ is
\begin{equation}
\boxed{\frac{\delta_{f_i}}{f_i} = \frac{\sigma_{f_i}^2}{f_i^2} \left(\sum_{j} \frac{f_j^2}{f_i^2} \frac{s_i^2 \sigma_{f_i}^2}{s_j^2 \sigma_{f_j}^2} \right)^{-1}}
\label{eqn:bias_joint_phot}
\end{equation}
where the sum over $j$ is taken over all bands used in the fit. If all bands have the same PSF size $s$, then all bands have the same fractional flux bias of $(\sum_j {\rm SNR}_j^2)^{-1}$.

We test our analytic predictions for simultaneous fitting by creating a set of two-band simulated point-source images with the same parameters as those used to make Figure \ref{fig:forced_photometry_bias} for forced photometry. Figure \ref{fig:simultaneous_fit_bias} shows that the fractional bias in both bands depends on the combined SNR, as predicted.

\section{Application}
\label{sec:real}

While the results discussed in the previous sections are present in our simulations, we now turn our attention to real-world datasets to demonstrate that these effects are likely present in most datasets currently used in the astronomy community.

\subsection{HSC SynPipe}
\label{subsec:hsc}

We first investigate whether this effect is present in more realistic mock catalogs processed by real pipelines. In particular, we use simulated data from Hyper Suprime-Came Subaru Strategic Program (HSC-SSP; \citealt{2018PASJ...70S...4A}) Synthetic Object Pipeline \citep[SynPipe;][]{2018PASJ...70S...6H}. In brief, SynPipe injects fake objects into real images which are then processed by the HSC-SSP Pipeline \citep{2018PASJ...70S...5B} to test the accuracy/precision of various aspects of the pipeline. These objects are drawn from a realistic color and magnitude distribution based in part on data from COSMOS, and so these mock tests represent fairly realistic realizations of the data seen by the HSC pipeline. See \cite{2018PASJ...70S...6H} for additional details.

\begin{figure}
\includegraphics[scale=0.19]{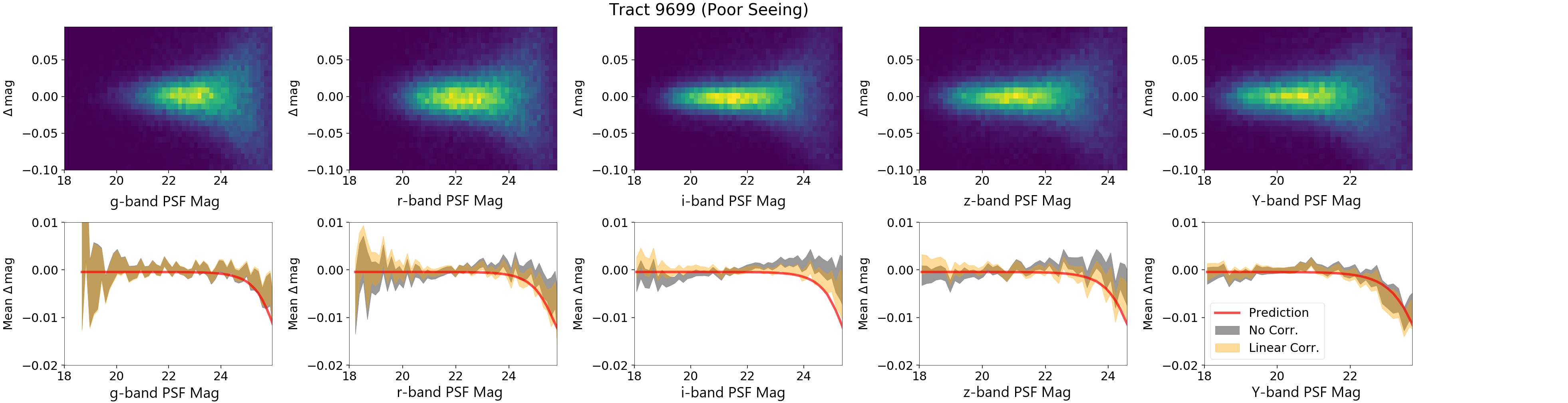}
\caption{As Figure \ref{fig:hsc_good}, but for a tract with poor seeing ($> 1 \arcsec$). While a number of additional systematic trends remain present in the data, the analytic prediction continues to provide a good fit to the observed trends.}
\label{fig:hsc_poor}
\end{figure}

We analyze the PSF magnitudes for artificial star tests from two tracts (8764 and 9699) with good/poor seeing, respectively, processed using the same SynPipe configuration presented in \cite{2018PASJ...70S...6H}. The corresponding magnitude offsets and the predicted analytic relations are shown in Figures \ref{fig:hsc_good} and \ref{fig:hsc_poor} for good seeing and poor seeing data, respectively. The magnitude offsets show good agreement with our model predictions, but include an additional systematic that is linear with magnitude. Investigating the source of this additional systematic is beyond the scope of this work\footnote{We have contacted many of the main authors of the pipeline so they are now aware of this systematic.}.

One crucial aspect of these results is that the SynPipe tests nominally represent \textit{forced} photometric extractions, with detection done in the $i$ band. However, our results are almost entirely consistent with \textit{unforced} photometry, where each band is derived separately. After some investigation, we find that this effect can be accounted for within the forced photometry algorithm used by the HSC pipeline, which effectively allows for limited ``re-centering'' in each band to improve the fit. Because the allowed range of positions is much larger than the relative positional uncertainties suggested by, e.g., $\sigma_x$ (\S \ref{subsec:ml_pos}) in most cases, this process effectively undoes the forcing effect described in \S \ref{subsec:phot_single}. This phenomenon -- where ``forced'' photometry from a particular pipeline is not quite what its namesake suggests -- highlights the importance of transparency when pipelines provide users with results for conducting detailed analysis.


\subsection{Stripe 82}
\label{subsec:s82}

To show that this bias appears in real data, we also look at SDSS catalogs of Stripe 82 \citep{2014ApJ...794..120A}. Stars that are low signal-to-noise in individual ``runs'' should have magnitudes that are biased high relative to their true values. While we do not have access to those true values, we approximate them using measurements taken from the combined images constructed from all the runs, which give much higher SNR measurements (with negligible bias) relative to the individual runs. We expect stars to be brighter, on average, in the individual run catalogs than in the stacked image catalog. Furthermore, each run and band will have a different bias, due to differences in seeing and sky brightness.

\begin{figure}
\includegraphics[scale=0.19, trim=20mm 15mm 30mm 5mm]{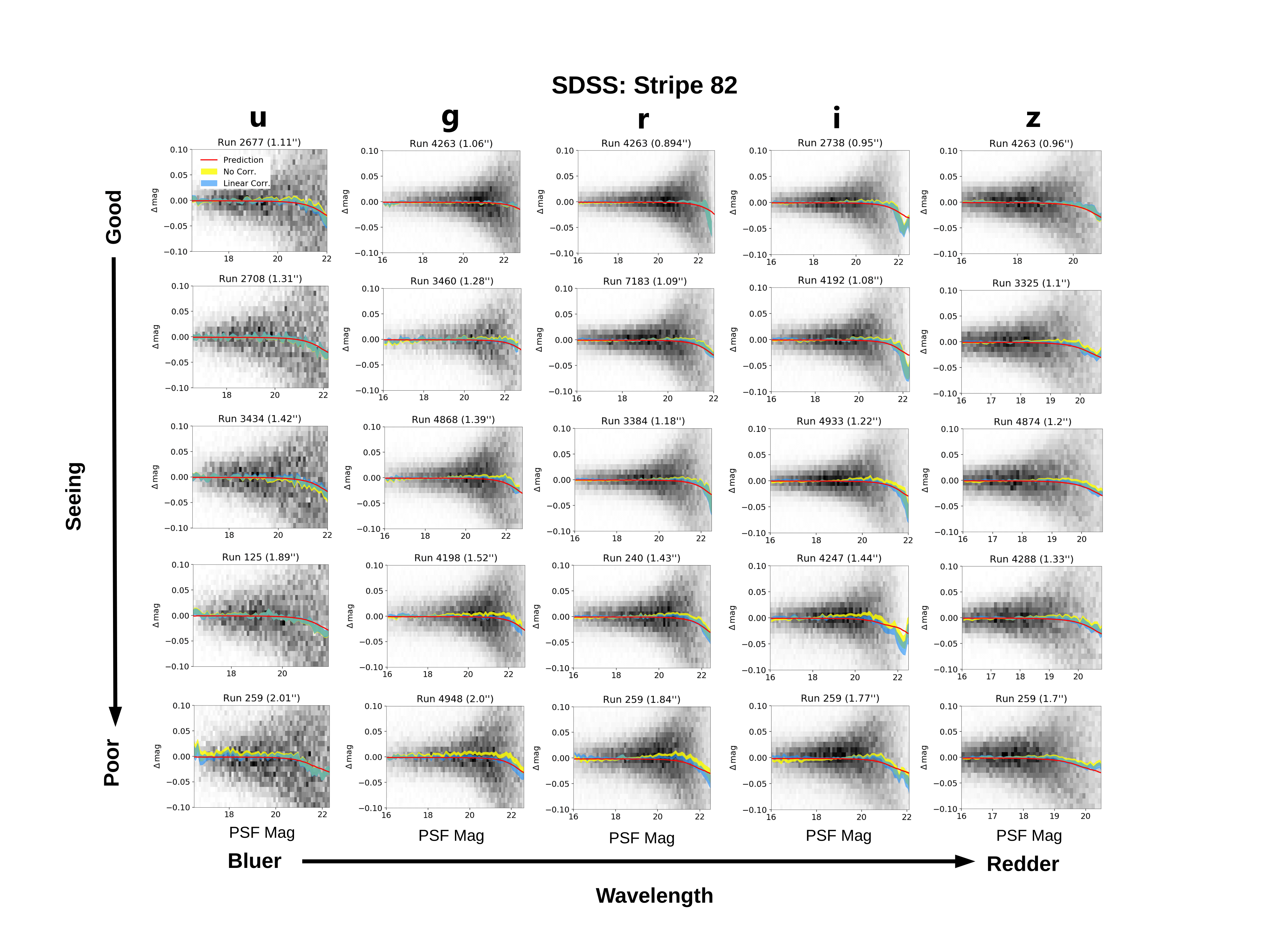}
\caption{The distribution of PSF magnitude offsets in the $ugriz$ bands for stars in SDSS Stripe 82 between individual runs and the measured values from the combined catalog. The general density is shown in grayscale, with the mean trends with and without a linear correction highlighted in blue and yellow, respectively, and our prediction (${\rm SNR}^{-2}$) in red. As Figure \ref{fig:hsc_good}, the mean magnitude offsets have been shifted to accommodate additional systematic offsets. While a number of additional systematic trends similar to those seen in the HSC SynPipe data (Figures \ref{fig:hsc_good} and \ref{fig:hsc_poor}) are present, there is good agreement between the predicted and observed offsets.}
\label{fig:stripe_82}
\end{figure}

Figure \ref{fig:stripe_82} shows the magnitude difference between the individual run catalogs and stacked image catalog for each band and over a range of seeing conditions. The faintest stars are biased brighter in the individual runs, in rough agreement with our predictions. It is interesting to note that the apparent systematic trends seen in these data mirror those in the HSC SynPipe tests, and that the photometry is also described as ``forced'' photometry from the SDSS pipeline. As the HSC pipeline is in part derived from the SDSS pipeline, these similarities bolster our suspicions that these results are most likely caused by the same algorithmic choices.

\section{Discussion}
\label{sec:disc}

\subsection{Aperture Photometry}
\label{subsec:phot_aperture}

We have shown that ML methods exhibit a generic bias when estimating the flux density of any particular isolated object from a given footprint. This bias causes flux density to be overestimated, and arises because fitting for an unknown object position does not treat noise fluctuations symmetrically. When the noise at the true position fluctuates low, the fit will be drawn away from the true position by nearby high noise fluctuations; conversely, when the noise at the true position fluctuates high, the fit will tend to remain near the true position. As is expected for any ML estimation bias, it is most pronounced at low SNR, scaling as ${\rm SNR}^{-2}$. This bias becomes worse as the models become more complex (as is the case for extended sources such as galaxies), and also biases colors when fitting across multiple bands with forced photometry.

Given these apparent drawbacks, some astronomers might wonder whether a return to aperture-based methods might present a compelling alternative. We want to offer a few arguments for why ML photometry should still be preferred and offer advice where aperture photometry might be more appropriate.

First, ML photometry performs better in the ideal case, where it is unbiased and follows the true error distribution, i.e. $f_\ml^* \sim \mathcal{N}(f^*, \sigma_f^2)$. By contrast, an aperture will ``miss'' part of an object's flux, leading to an \textit{underestimate} in \textit{all} cases. This correction is not expected to be the same for all sources unless the aperture is adaptively adjusted to match (a few times) the size of the PSF-convolved object, which is rarely the case. The ``aperture corrections'' involved to capture the total flux subsequently almost always serves as a dominant systematic hindering precise analyses.

Aperture photometry might also not eliminate the ``centering bias'' described in this work. Since an aperture also requires a position $(x, y)$ to be centered on, determining a central position for the aperture will likely be subject to the same types of biases as the ML case (\S\ref{subsec:bias_gen}).\footnote{Indeed, apertures are often centered \textit{using} positions determined from either model-fitting approaches or various simplistic heuristics (e.g., ``peak hunting'').} 
These expected centering offsets will result in variable amounts of flux being excluded from the aperture, likely biasing aperture photometry to a similar extent as ML photometry. Unlike in the ML case where these biases can be studied using statistical methods, however, apertures by nature make such studies much more difficult.

The derived errors from aperture photometry are also generally larger than those from ML photometry. In the ML case, we showed in \S\ref{sec:preliminaries} that the error for a point source is $\sigma_f^2 = \neff \sigma^2$, where $\neff$ is the effective area of the PSF. For a Gaussian PSF with a standard deviation of $s$ pixels, this gives $A^G_{\rm psf} = \pi (2s)^2$. Since aperture photometry just sums all pixels within a given aperture, the equivalent error for an aperture with radius of $r=2s$ pixels is just $\sigma_f^2 = \pi r^2 \sigma^2 = \pi (2s)^2 \sigma^2$. Any aperture larger than ``2-sigma'' then has errors that are strictly larger than those estimated from ML photometry, and even this $2s$ aperture excludes roughly 5\% of the flux, requiring a significant aperture correction.

Aperture photometry is also inherently \textit{unstable} as the aperture increases. While increasing the size of the aperture ensures a greater amount of the total flux is captured, it also increases the variance proportional to the size. While the SNR from ML photometry strictly \textit{improves} as more data is added (see \S\ref{subsubsec:back_dec}), the SNR from aperture photometry strictly \textit{degrades} (ignoring aperture corrections). 

Finally, aperture photometry is unable to integrate information across multiple bands/images. As discussed in \S\ref{subsec:phot_unforced}, simultaneously fitting a single model across multiple images strictly improves the SNR and reduces the effective bias. Because apertures assume no model, they are unable to improve their SNR across multiple bands. While this comparison appears to be irrelevant in the examples shown in \S\ref{sec:real} which all exhibit tendencies equivalent to single-band unforced photometry, it will likely become more relevant in future survey pipelines.

\begin{figure}
\includegraphics[scale=0.23]{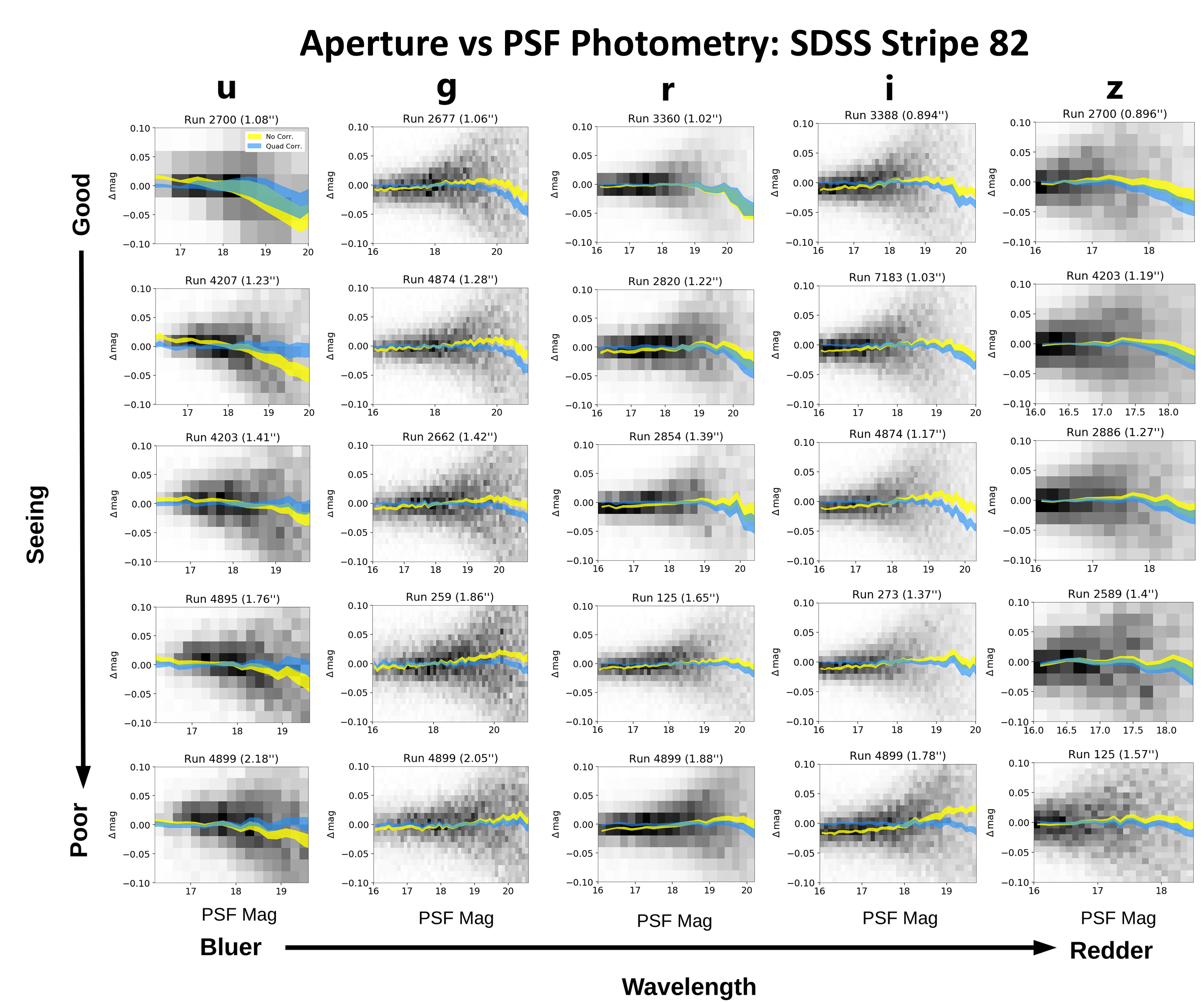}
\caption{As Figure \ref{fig:stripe_82}, but now showing the differences between PSF and aperture magnitudes for the same stars between individual runs versus the measured PSF magnitudes from the combined catalog. The general density is shown in grayscale, with the mean trends with and without a quadratic correction highlighted in blue and yellow, respectively. As Figure \ref{fig:hsc_good}, the mean magnitude offsets have been shifted to accommodate additional systematic offsets. There are large differences and additional systematic trends similar to those seen in the HSC SynPipe data (Figures \ref{fig:hsc_good} and \ref{fig:hsc_poor}) at bright magnitudes that are likely due to issues with background estimation and aperture flux corrections. Once these trends are removed, we see evidence for a bias that scales as SNR$^{-2}$ appearing at much brighter magnitudes than in the PSF magnitude-only comparison (Figure \ref{fig:stripe_82}) due to the substantially larger errors associated with the aperture photometry measurements.}
\label{fig:aper}
\end{figure}

We illustrate the magnitude of the above effects in Figure \ref{fig:aper}, which shows the magnitude difference between PSF and aperture (\texttt{aper6}) measurements from the same sources in the individual run catalogs as a function of the PSF magnitude measured from the stacked image catalog for each band and over a range of seeing conditions. There are substantial differences between the two measurements even at bright magnitudes that also show complex dependencies as a function of seeing. These most likely arise from the increasing sensitivity of the aperture to issues like background mis-estimation and incorrect aperture corrections. Once these effects are removed, we see behavior suggestive of a SNR$^{-2}$ bias. Compared to PSF photometry (see Figure \ref{fig:stripe_82}), this bias becomes significant at much brighter magnitudes due to substantially larger aperture flux errors.

Ultimately, aperture photometry is appealing because it is so simple: it assumes no model and is straightforward to apply to almost any isolated object. While this leads to many of the drawbacks mentioned above and shown in Figure \ref{fig:aper}, it can also be desirable in cases where modeling complex sources can be difficult and/or the systematics involved limit the effective SNR of an object below that achievable with ML photometry. It thus serves a valuable purpose in cases where a model for the PSF and/or source cannot be cleanly determined \textit{and} the source is relatively isolated. Our results suggest that it should only be used judiciously in most other cases.

\subsection{Stacked Catalogs}
\label{subsec:phot_stackcat}

\begin{figure}
\includegraphics[scale=0.19]{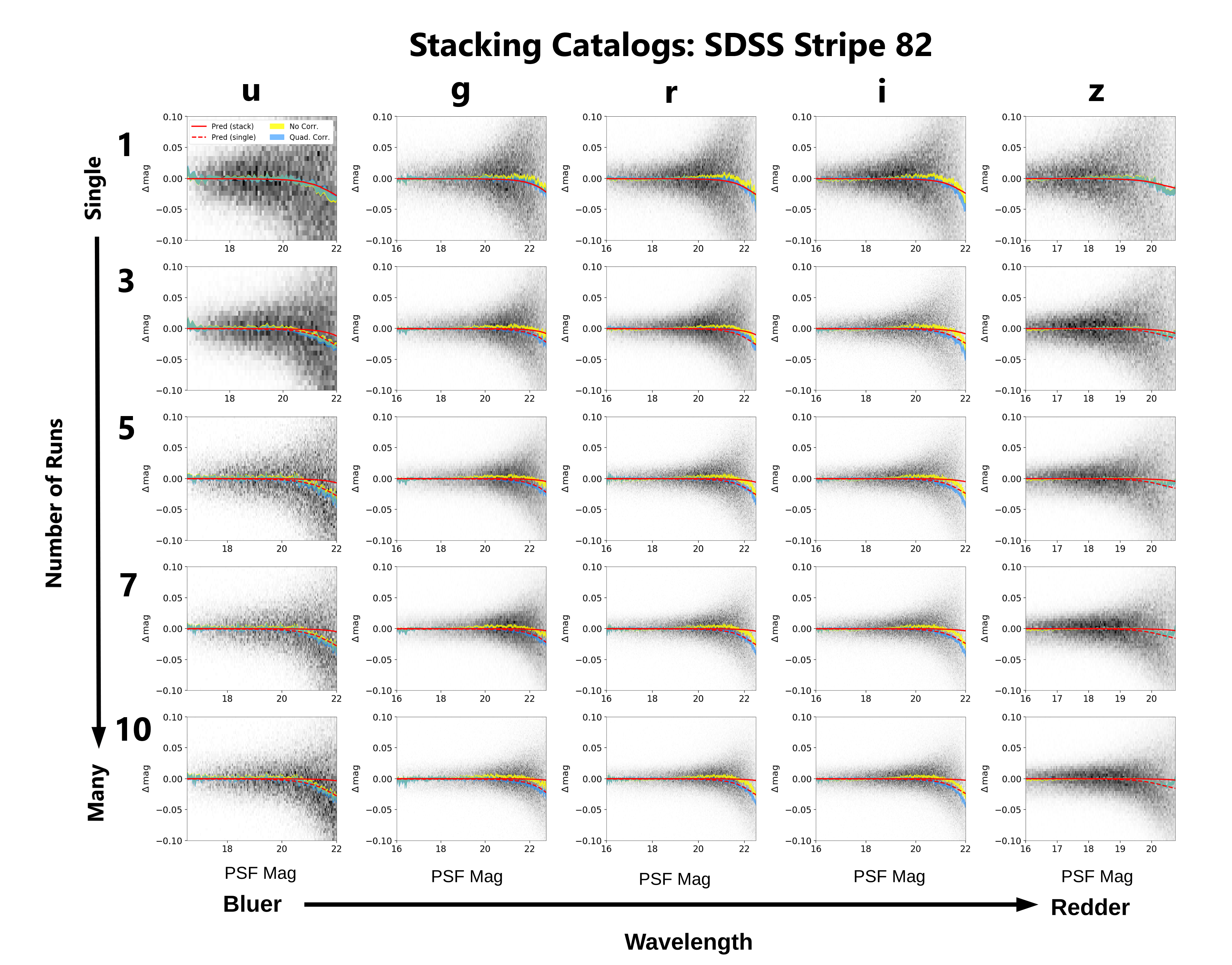}
\caption{As Figure \ref{fig:stripe_82}, but now for the case where we have constructed ``deeper'' catalogs by averaging objects detected across individual runs \textit{on the catalog level} rather than co-adding the images and we have applied a \textit{quadratic} rather than linear correction for catalog-related biases. The general density is shown in grayscale, with the mean trends with and without the quadratic correction highlighted in blue and yellow, respectively. Predictions of the bias using the combined SNR of the images are in red solid lines, and our predictions using the SNR of a single image are in red dashed lines.} When dealing with a single run (top row), the predicted bias (which goes as ${\rm SNR}^{-2}$) is a good fit to the data. Since this bias is the same across all runs, stacking observations on the catalog level (bottom rows) does not reduce this bias compared to the single-run case (red dashed lines) even though the nominal errors are smaller. Assuming that the bias decreases with the stack's flux uncertainty leads to an underestimate of the bias (solid red lines), which can quickly become larger than the estimated errors. Stacking on the catalog-level can also compound other systematic effects introduced when modeling the images from each run.
\label{fig:s82_stack}
\end{figure}

One direct corollary of our results is that users must be extremely careful what exactly ``stacking'' means when constructing catalogs and estimating photometry from sources. This is important because stacking can occur at multiple levels, ranging from the images themselves to the catalogs produced from them. In the former case, where images are combined \textit{before} they are processed through a pipeline, modeling the results is somewhat equivalent to the simultaneous fitting approach discussed in \S\ref{subsec:phot_unforced}. Assuming that each image has roughly the same error $\sigma_f$, the effective error from $N$ images is expected to decrease to $\sigma_f/\sqrt{N}$ and the bias to decrease accordingly. Stacking on the image level thus reduces both the error and the bias.

In the case where stacking is done on the \textit{catalog} level, however, each observation will be biased, with a mean of $f_\ml = f^*(1+{\rm SNR}^{-2})$ and error of $\sigma_f$. Averaging the results over many identical catalogs then will reduce the error to $\sigma_f/\sqrt{N}$, but \textit{will not} decrease the bias in any meaningful way. This implies that any measurement constructed from a stacked catalog may have systematic biases that far exceed the quoted statistical uncertainties. 

Put another way, making catalogs from a series of images and then averaging the measurements across catalogs will not remove the flux density bias, because each catalog is individually biased. If inverse variance weighting is used and all images have the same PSF size, the fractional bias of the average will then be the reciprocal of the \textit{average} ${\rm SNR}^2$. In contrast, the fractional bias from image stacking or simultaneous fitting is the reciprocal of the \textit{total} ${\rm SNR}^2$, allowing multiple images of comparable SNR to drive down the bias. If catalogs are to be averaged, each catalog should first be \textit{individually} debiased so that the average flux across catalogs is also unbiased. This procedure increases the variance in each catalog's flux since the exact bias for each is not known (the bias-variance trade-off; see \S\ref{sec:bias}); however, this increase in variance is similar to that incurred by debiasing the flux measured from a stacked image or simultaneous fit.

For example, \cite{Budavari2017} propose detecting faint sources by stacking at the catalog level rather than the image level. This approach requires making catalogs with low SNR thresholds from each image, which is precisely the regime where the ML flux bias is most important. The catalogs made from each individual image should be debiased before being stacked.

We illustrate this effect in Figure \ref{fig:s82_stack} by constructing ``stacked catalogs''\footnote{We ``stack'' our observations in flux density (linear) space to compute the weighted arithmetic mean. Note that stacking in magnitude (logarithmic) space (which computes the weighted geometric mean) introduces additional biases.} using the same SDSS Stripe 82 data used to generate Figure \ref{fig:stripe_82}. As expected, the bias remains unchanged regardless of the number of runs used to generate the stack even as the estimated errors (and thus the bias we would \textit{predict} from the stacked catalogs) decreases substantially.

\subsection{Bayesian Inference}

The maximum likelihood bias we described can be ameliorated by using Bayesian inference. We show in Appendix \ref{ap:flux_mean} that marginalizing over the possible positions of the source cancels out the ML bias. Bayesian methods using Markov chain Monte Carlo sampling are starting to be implemented in astronomy. The widely-used Tractor code \citep{Lang2016} returns ML measurements but also allows users to sample the posterior distribution of source parameters. Probabilistic cataloging \citep{Brewer2013,Portillo2017} allows users to sample from the posterior in catalog space, yields an ensemble of catalogs that may vary in the number of sources identified. This catalog ensemble is useful in crowded fields to capture uncertainties in source identification and deblending. The computational cost required to sample from a posterior rather than reporting ML parameters can be daunting. As such, it is likely the case that ML methods will continue to be applied in both the near and intermediate terms in modern astronomical image processing pipelines.

\subsection{Galaxies}
\label{subsec:gal}

\edit1{
In much of modern astronomy users are interested in obtaining photometric measurements for resolved, extended sources such as galaxies in addition to unresolved point sources. As outlined in Appendix \ref{ap:bias}, our results can be generalized to a $p$-parameter object model to give
\begin{equation}
f_\ml^* \approx f_\ml \left[ 1 - \frac{X_{p-1}^2}{2}\frac{\sigma^2_{f_\ml}}{f_\ml^2} \right]
\end{equation}
where $X^2_{p-1} \sim \chi^2_{p-1}$ is a chi-square random variable with $p-1$ degrees of freedom, $\sigma^2_{f_\ml}$ is the ``true'' error estimate that includes the covariances from the other object parameters as well as the background, and the $p-1$ comes from the fact that we are excluding the flux density $f$. This gives
\begin{equation}
\boxed{
\frac{\delta_{f_\ml}}{f_\ml} \approx \frac{p-1}{2}\frac{\sigma^2_{f_\ml}}{f_\ml^2}, \quad
\frac{\mathbb{V}[f_\ml^*]}{\sigma_{f_\ml}^2} \approx \frac{p^2 - 4p + 7}{4} \frac{\sigma^2_{f_\ml}}{f_\ml^2}
}
\label{eqn:bias_par_gal}
\end{equation}
Galaxies thus increase the bias and underestimate the variance both by including more free parameters (increasing $p$) as well as by introducing covariances among them (increasing $\sigma^2_{f_\ml}$).
}

\edit1{
When modeling galaxies using single-component models (e.g. Sersic profiles), astronomers typically introduce anywhere from 1-4 additional parameters beyond just position $(x,y)$ and flux density $f$. These at a minimum often include parameters to model the physical size (e.g., effective radius $R_e$), elongated shapes/projection effects (e.g., axis ratio $b/a$ and position angle $\phi$), and surface brightness profiles (e.g., scale index $n$). For multi-component models such as \texttt{cmodel} \citep{2004AJ....128..502A}, this can include up to 8 additional parameters.
}

\edit1{
Due to the additional parameters involved, we should expect the ML flux density estimates for galaxies (and other extended objects) to \textit{at least} double. In addition, because galaxy models can introduce potentially strong covariances between $f$ and other model parameters, we would expect $\sigma_f^2$ to increase even in the absence of any background modeling by \textit{at least} a factor of a few. 
These combined effects (larger effective image areas, more parameters, and stronger parameter covariances) imply that we expect biases arising purely from our ML approach to now be roughly 0.6\% at $20\sigma$, 2.5\% at $10\sigma$, and 10\% at $5\sigma$.
}

\begin{figure}
\includegraphics[scale=0.3]{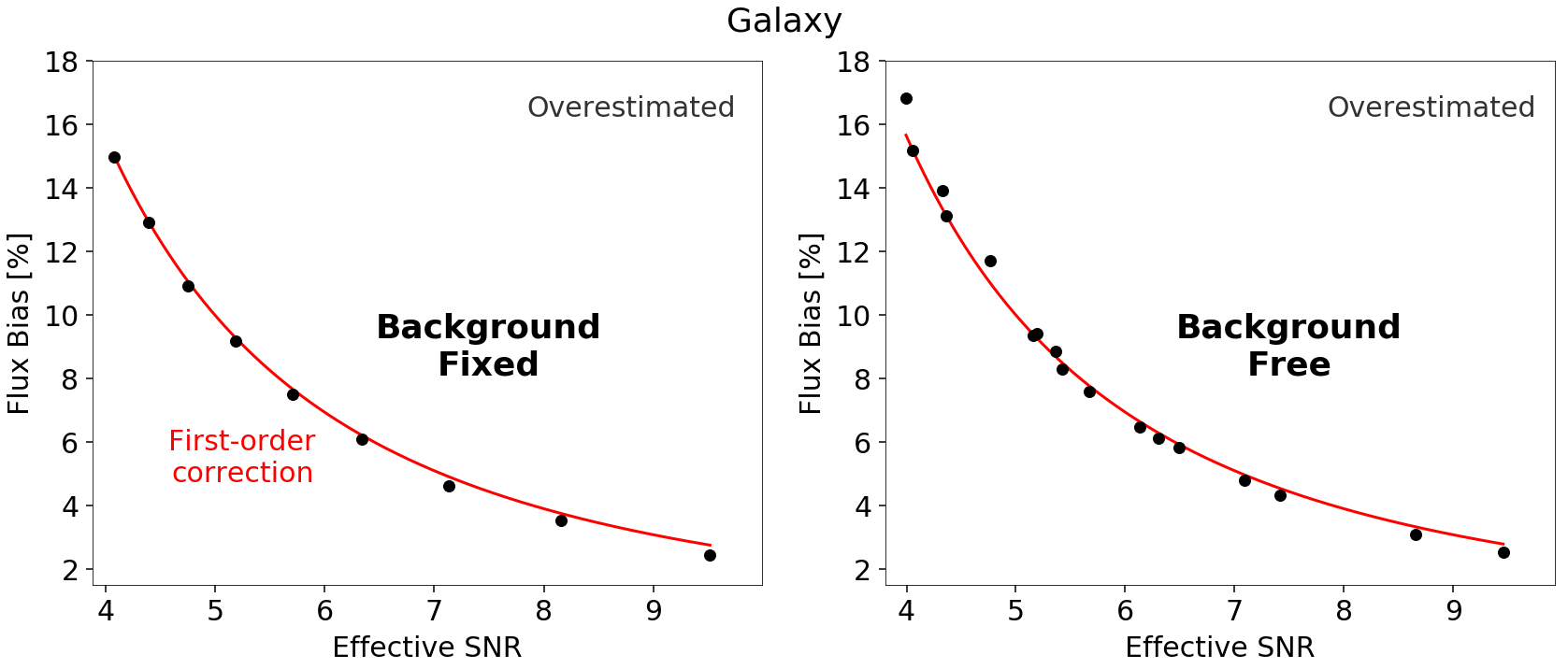}
\caption{The bias in the estimated maximum-likelihood flux density $f_\ml$ for a 2-D ``Gaussian galaxy'' assuming a fixed (left) and varying (right) background over several image sizes. The first-order analytic predictions (Appendices \ref{ap:bias} and \ref{ap:gaussgal}) are shown in red. At fixed SNR, $f_\ml$ is biased higher than in the point-source case from Figure \ref{fig:star_fixed_back} because the shape parameters can ``absorb'' additional noise in the image, further increasing the log-likelihood and estimated flux. To leading order, the fractional flux bias is $5/2 \times {\rm SNR}^{-2}$.}
\label{fig:galaxy_bias}
\end{figure}

\edit1{
We created a set of simulated images of a 2D Gaussian galaxy and run maximum-likelihood photometry on them, with the standard deviations $s_1$ and $s_2$ and position angle $\phi$ left free parameters. Our simulated images have a circular Gaussian galaxy of $s_1=s_2=1\;\rm{pixel}$ with a circular Gaussian PSF of $\sigma=2\;\rm{pixels}$. Again, each pixel has iid Gaussian noise. For the fixed-background case, we simulate sources of nine fluxes ranging from $4.3\sigma$ to $10.0\sigma$, evenly spaced in $1/SNR$. For the free background case, we consider the same fluxes in the previously used five image sizes (11, 13, 15, 23, and 101 pixels). Again, we create 100,000 different simulated images for each configuration. Figure \ref{fig:galaxy_bias} shows that the mean flux bias from the simulated galaxy images agree well with theoretical predictions outlined in Appendix \ref{ap:gaussgal}
in both the fixed background and free background case.
}

\edit1{
Although we have focused on flux density estimates, we note that \citet{2012MNRAS.425.1951R} have also examined a similar bias in derived galaxy shapes and sizes.\footnote{They also derive a bias in the estimated flux density which agrees with our results.} We encourage interested readers to examine their work for additional details.
}

\edit1{Unfortunately, there are many other complications when dealing with extended sources besides this expected ML flux bias. In particular, the additional complexity of the 2D light profiles of extended sources can lead to substantial \textit{model mismatch} between the simple generative model we assume and the actual data. They can also be much more sensitive to mis-estimation of and spatial variation in the PSF. These issues make it difficult to accurately predict systematic biases in galaxy photometry in real datasets without the use of realistic fake-object pipelines that can perform end-to-end simulations of complex extended sources. As such, we expect our results to serve as a rough guide as to the expected level and behavior of the flux bias \textit{in the best possible case}.}

\section{Conclusion}
\label{sec:conc}

In this paper, we study a photometric bias that arises from the maximum-likelihood (ML) estimator in model fitting photometry. It arises (in part) because the ML estimate can ``soak up'' a small amount of noise such that the fitted position is drawn away from the true position. This leads to an overestimate of the flux density, with a bias scaling with the inverse signal-to-noise ratio (${\rm SNR}^{-2}$) and the number of free parameters in the model. For example, it is 1\% for a $10\sigma$ point-source and 2.5\% for a $10\sigma$ 2-D Gaussian galaxy.

While this leads to an overestimate in the detection band, because the derived position is offset from the true position the flux density will be underestimated by the same ${\rm SNR}^{-2}$ in any other bands where the position is forced to the same value. This can double the effective bias in derived colors. By contrast, when all bands are modeled simultaneously, all bands are biased high, but less so than if they had been fit individually. In the case where all of the PSFs are the same size, this bias goes as $(\sum {\rm SNR}^2)^{-1}$. If an object's position is already known to great precision (for example, from a deeper or higher-resolution dataset), then forced photometry using this fixed position also does not suffer this bias. Methods that consider the distribution of possible positions, like Bayesian inference, do not exhibit this bias (see Appendix \ref{ap:flux_mean}).

We provide formulae that can be used in a variety of situations to calculate this bias and thus correct for it:
\begin{itemize}
    \item Single-band point source photometry (Equation \ref{eqn:bias_point_source})
    \item Multi-band forced point source photometry (Equation \ref{eqn:bias_forced_phot})
    \item Multi-band joint point source photometry (Equation \ref{eqn:bias_joint_phot})
    \item \edit1{Parameterized extended source (single-band model)} (Equation \ref{eqn:bias_par_gal})
\end{itemize}

We then show that this bias is likely common in many astronomical datasets using both mock HSC-like data and real SDSS data. The results further illustrate the importance of pipelines being transparent about the exact algorithmic implementation, since both tests are consistent with unforced photometry even though the data have been extracted using ``forced'' methods.

Although maximum-likelihood estimators may be biased, we still strongly encourage using them over simpler aperture-based methods in most cases. While apertures are appealing because of their simplicity, they require difficult-to-model aperture corrections to account for missing flux and likely exhibit similar biases due to offsets in aperture centers relative to the true positions of objects. Apertures also cannot effectively incorporate information across multiple bands, which can substantially reduce any relevant biases by improving the effective signal-to-noise ratio.

Though we have shown derivations in a simplified case, with iid Gaussian noise, Gaussian PSFs, and Gaussian galaxy profiles, this bias is generic to maximum-likelihood photometry estimation and would still arise if these assumptions were relaxed. This analysis could be repeated under different assumptions, like Poissonian noise, real PSFs, and Sersic galaxy profiles. While calculating the relevant corrections will likely be more involved in these more realistic cases, they are likely still tractable through the bias tensor formalism (See Appendix \ref{ap:bias_tensor}) or through numerical simulations.

\acknowledgments

\edit1{The authors would like to thank the anonymous referee for providing comments that improved the quality of this work.} The authors would also like to thank Jim Bosch, Charlie Conroy, Daniel Eisenstein, Song Huang, {\v Z}eljko Ivezi\'c, Ben Johnson, Douglas Scott, Eddie Schlafly, Sandro Tacchella, and Catherine Zucker for discussion and feedback that helped to improve the quality of this work. JSS is eternally grateful to Rebecca Bleich for her patience and support.

JSS is supported by the National Science Foundation Graduate Research Fellowship Program. SKNP acknowledges support from the DIRAC Institute in the Department of Astronomy at the University of Washington. The DIRAC Institute is supported through generous gifts from the Charles and Lisa Simonyi Fund for Arts and Sciences, and the Washington Research Foundation.

The Hyper Suprime-Cam (HSC) collaboration includes the astronomical communities of Japan and Taiwan, and Princeton University. The HSC instrumentation and software were developed by National Astronomical Observatory of Japan (NAOJ), Kavli Institute for the Physics and Mathematics of the Universe (Kavli IPMU), University of Tokyo, High Energy Accelerator Research Organization (KEK) in Japan, Academia Sinica Institute for Astronomy and Astrophysics (ASIAA) in Taiwan, and Princeton University in the United States. Funding was contributed by the FIRST program from Japanese Cabinet Office; Ministry of Education, Culture, Sports, Science and Technology (MEXT); Japan Society for the Promotion of Science (JSPS); Japan Science and Technology Agency (JST); Toray Science Foundation; NAOJ; Kavli IPMU; KEK; ASIAA; and Princeton University.

This paper makes use of data products derived from software developed for the Large Synoptic Survey Telescope (LSST). We thank the LSST Project for making their code available as free software at \url{http://dm.lsstcorp.org}.

Funding for the Sloan Digital Sky Survey IV has been provided by the Alfred P. Sloan Foundation, the U.S. Department of Energy Office of Science, and the Participating Institutions. SDSS acknowledges support and resources from the Center for High-Performance Computing at the University of Utah. The SDSS web site is \url{www.sdss.org}. 

SDSS is managed by the Astrophysical Research Consortium for the Participating Institutions of the SDSS Collaboration including the Brazilian Participation Group, the Carnegie Institution for Science, Carnegie Mellon University, the Chilean Participation Group, the French Participation Group, Harvard-Smithsonian Center for Astrophysics, Instituto de Astrof\'isica de Canarias, The Johns Hopkins University, Kavli Institute for the Physics and Mathematics of the Universe (IPMU) / University of Tokyo, the Korean Participation Group, Lawrence Berkeley National Laboratory, Leibniz Institut f\"ur Astrophysik Potsdam (AIP), Max-Planck-Institut f\"ur Astronomie (MPIA Heidelberg), Max-Planck-Institut f\"ur Astrophysik (MPA Garching), Max-Planck-Institut f\"ur Extraterrestrische Physik (MPE), National Astronomical Observatories of China, New Mexico State University, New York University, University of Notre Dame, Observat\'orio Nacional / MCTI, The Ohio State University, Pennsylvania State University, Shanghai Astronomical Observatory, United Kingdom Participation Group, Universidad Nacional Aut\'onoma de M\'exico, University of Arizona, University of Colorado Boulder, University of Oxford, University of Portsmouth, University of Utah, University of Virginia, University of Washington, University of Wisconsin, Vanderbilt University, and Yale University.

\bibliography{references}
\bibliographystyle{aasjournal}

\appendix

\section{Results with Correlated Noise}
\label{ap:gen_res}

Assume the noise in our footprint to be Normally distributed such that
\begin{equation}
\hat{\bflux} \sim \mathcal{N}(\bflux^*, \cov)
\end{equation}
with mean vector $\bflux^*$ and covariance matrix $\cov$. The corresponding log-likelihood for a point source within our footprint is
\begin{equation}
\ln \mathcal{L}(x,y,f,\bflux) = -\frac{1}{2} \ln(\det(2\pi \cov)) -\frac{1}{2} (\hat{\flux} - \bflux - f\psf_{x.y})^T \cov^{-1} (\hat{\flux} - \bflux - f\psf_{x,y})
\end{equation}
where $\det$ is the determinant (i.e. the dimension-independent analog of area/volume).

\subsection{Flux Density}
\label{subap:gen_res_flux}

At the ML flux $f_{\ml}(x,y,\bflux)$ the derivative with respect to $f$ is zero such that
\begin{equation}
\partial_f\ln\mathcal{L}(x,y,f_\ml,\mathbf{b}) = (\hat{\flux} - \bflux)^\T \cov^{-1} \psf_{x,y}  - f_\ml \psf_{x,y}^\T \cov^{-1} \psf_{x,y} = 0
\end{equation}
which yields
\begin{equation}
\boxed{f_\ml(x,y,\bflux) = \frac{(\hat{\flux} - \bflux)^\T \cov^{-1} \psf_{x,y}}{\psf_{x,y}^\T \cov^{-1} \psf_{x,y}}}
\end{equation}
The naive estimate (see \S\ref{subsec:ml_flux}) of the error/uncertainty $\tilde{\sigma}_f(x,y)$ is then
\begin{equation}
\boxed{\tilde{\sigma}_f^2(x,y) = \left(\psf_{x,y}^\T \cov^{-1} \psf_{x,y}\right)^{-1}}
\end{equation}

\subsection{Position}
\label{subap:gen_res_pos}

The maximum-likelihood positions $(x_\ml,y_\ml)$ can likewise be defined via
\begin{equation}
\partial_x \ln\mathcal{L}(x_\ml,y,f,\mathbf{b}) = f \left( (\hat{\flux} - \bflux) - f \psf_{x_\ml,y} \right)^\T \cov^{-1} \partial_x \psf_{x_\ml,y} = 0
\end{equation}
with a naive error/uncertainty of
\begin{align}
\tilde{\sigma}_x^2(x_\ml,y,f,\bflux) &= \frac{1}{f^2} \left( \partial_x \psf_{x_\ml,y}^\T \cov^{-1} \partial_x\psf_{x_\ml,y} - \frac{1}{f} \left( (\hat{\flux} - \bflux) - f \psf_{x_\ml,y} \right)^\T \cov^{-1} \partial^2_x\psf_{x_\ml,y} \right)^{-1} \nonumber \\ 
&\approx \frac{1}{f^2} \left(\partial_x\psf_{x_\ml,y}^\T \cov^{-1} \partial_x\psf_{x_\ml,y}\right)^{-1}
\end{align}

\subsection{Background}
\label{subap:gen_res_back}

The maximum-likelihood background $\bflux_\ml$ can likewise be defined using
\begin{equation}
\partial_{\bflux}\ln\mathcal{L}(x,y,f,\mathbf{b}_\ml) = \cov^{-1} (\hat{\flux} - f\psf(x,y) - \bflux_\ml) = 0
\end{equation}
which gives
\begin{equation}
\bflux_\ml(x,y,f) = \hat{\flux} - f\psf_{x,y} \label{eqn:back_gen}
\end{equation}
The associated naive errors then are
\begin{equation}
\tilde{\cov}_\bflux = \cov
\end{equation}

This result shouldn't be entirely surprising. In \S\ref{subsec:ml_back}, we noted that the maximum-likelihood background is just the mean residual between the model $f\psf_{x,y}$ and the data $\hat{\flux}$ in our given $\xdim \times \ydim$ footprint. In the iid case where we have assumed a fixed value $b$ across the footprint, we therefore take the average over all the pixels. In the case where every pixel has a separate possible value $b_{\ml,i}$ for the background, however, this leads to averaging done on a per-pixel basis for $\bflux_\ml$. Since fitting the background separately in every pixel always over-fits the data by construction, we also derive results for the more realistic case where $\bflux_{\bparams} \equiv \bflux(\bparams)$ is actually a function of a $\bdim$ nuisance parameters $\bparams$ across the footprint in Appendix \ref{ap:back_gen}.

\subsection{Bias}
\label{subap:gen_res_bias}

As in \S\ref{sec:bias}, we can rewrite our likelihood in terms of random variable notation such that
\begin{equation}
\hat{\flux} = f^*\psf(x^*, y^*) + \bflux^* + \cov^{1/2}\normvec \sim \mathcal{N}(f^*\psf(x^*, y^*) + \bflux^*, \cov) 
\end{equation}
where each $Z_i \sim \mathcal{N}(0, 1)$ is an iid random variable drawn from the standard Normal distribution and $\cov^{1/2}$ is the (symmetric) square root of the covariance matrix. The likelihood at the true position $(x^*, y^*)$ and background $b^*$ then is
\begin{equation}
\ln \mathcal{L}(x^*,y^*,f,\bflux^*) = -\frac{1}{2} \ln(\det(2\pi \cov)) -\frac{1}{2} ((f^* - f)\psf_{x^*,y^*} + \cov^{1/2} \normvec)^\T \cov^{-1} ((f^* - f)\psf_{x^*,y^*} + \cov^{1/2} \normvec)
\end{equation}
This gives a ML flux density estimate of
\begin{equation}
f_\ml(x^*,y^*,\bflux^*) = f^* + \frac{\psf_{x^*,y^*}^\T \cov^{-1/2} \normvec}{\psf_{x^*,y^*}^\T \cov^{-1} \psf_{x^*, y^*}}
\end{equation}
Since
\begin{equation}
\frac{\psf_{x^*,y^*}^\T \cov^{-1/2} \normvec}{\psf_{x^*,y^*}^\T \cov^{-1} \psf_{x^*,y^*}} \sim \mathcal{N}\left(0, \frac{\psf_{x^*,y^*}^\T\cov^{-1}\psf_{x^*,y^*}}{\left(\psf_{x^*,y^*}^\T\cov^{-1}\psf_{x^*,y^*}\right)^2} = \tilde{\sigma}_f^2(x^*,y^*)\right)
\end{equation}
we then recover
\begin{equation}
\boxed{f_{\ml}(x^*,y^*,b^*) \sim \mathcal{N}(f^*, \sigma_f^2(x^*,y^*))}
\end{equation}
following \S\ref{subsec:bias_ideal}.

\section{Results with Correlated Parameter Uncertainties}
\label{ap:error}

\edit1{
The naive uncertainty estimates $\tilde{\sigma}_f$ derived in \S\ref{sec:preliminaries} and utilized in most of \S\ref{sec:bias} only are equal to the \textit{true} uncertainties $\sigma_f$ if there is no covariance between the $p \times 1$ parameter vector $\params$ that comprises our model. In general, we can assume the likelihood is approximately multivariate normal so that
\begin{equation}
\params_{\rm true} \sim \mathcal{N}(\params_\ml, \cov_{\params}(\params_\ml))
\end{equation}
where $\theta_\ml$ is the ML estimator. Following the discussion in \S\ref{sec:preliminaries} (see also Appendix \ref{ap:bias}), we can estimate the covariance around the ML solution as
\begin{equation}
\cov_{\params}(\params_\ml) \approx -\mathbb{E}_{\mathbf{D}}\left[\partial^2_{\params} \ln \mathcal{L}(\params_\ml) | \params_\ml \right]^{-1}
\equiv \left(\fisher_{\params}(\params_\ml)\right)^{-1} = -\left(\partial^2_{\params} \ln \mathcal{L}(\params_\ml)\right)^{-1}
\end{equation}
Here, $\mathbb{E}_{\mathbf{D}}[\cdot|\params_\ml]$ is the expectation value (i.e. mean) with respect to (random realizations of) the data $\mathbf{D}$ with $\params_\ml$ fixed. $\partial_{\params} \ln \mathcal{L}$ represents the $p \times 1$ Jacobian vector with respect to $\params$ with elements $(\partial_{\params} \ln \mathcal{L})_i = \partial_{\theta_i} \ln \mathcal{L}$, $\partial^2_{\params} \ln \mathcal{L}$ represents the $p \times p$ Hessian matrix whose elements are comprised of the second-order derivatives of the log-likelihood $(\partial_{\params} \ln \mathcal{L})_{ij} = \partial_{\theta_i} \partial_{\theta_j} \ln \mathcal{L}$, and $\fisher_{\params}(\params_\ml)$ is the \textbf{Fisher Information Matrix (FIM)} evaluated at $\params_\ml$. The final equality, where we remove the expectation value and equate the FIM with the inverse of the Hessian, follows from our assumption that the likelihood is multivariate normal.
}

\edit1{
With this more general result, we see that the results derived in \S\ref{sec:preliminaries} for the associated marginal uncertainties on each parameter $\tilde{\sigma}$ only are equal to the true marginal uncertainties $\sigma$ when the FIM is diagonal. When the off-diagonal elements of the FIM (i.e. mixed derivatives) are non-zero, the actual marginal uncertainties along the diagonal of $C_{\params}(\params_\ml)$ for the ML solution are larger. In this section, we examine the special $2 \times 2$ case involving just the flux density $f$ and one other parameter $\param_i$ for illustrative purposes.
}

\subsection{Error Underestimation in Two-Parameter Models}
\label{subsec:fisher_2x2}

\edit1{
In the special case where the FIM is $2 \times 2$, the inverse has a simple analytic form of
\begin{align}
C_{\params}(\params_\ml) &= \left[\fisher_{\params}(\params_\ml)\right]^{-1} =
- \begin{bmatrix}
\partial^2_{\param_i} \ln \mathcal{L}(\params_\ml) & \partial_{\param_i}\partial_{\param_j} \ln \mathcal{L}(\params_\ml) \\
\partial_{\param_i}\partial_{\param_j} \ln \mathcal{L}(\params_\ml) & \partial^2_{\param_j} \ln \mathcal{L}(\params_\ml) \\
\end{bmatrix}^{-1} \nonumber \\
&= \frac{1}{\det(\fisher_{\params}(\params_\ml))} 
\begin{bmatrix}
-\partial^2_{\param_j} \ln \mathcal{L}(\params_\ml) & \partial_{\param_i}\partial_{\param_j} \ln \mathcal{L}(\params_\ml) \\
\partial_{\param_i}\partial_{\param_j} \ln \mathcal{L}(\params_\ml) & -\partial^2_{\param_i} \ln \mathcal{L}(\params_\ml) \\
\end{bmatrix}
\end{align}
where the determinant is
\begin{equation}
\det(\fisher_{\params}(\params_\ml)) = \partial_{\param_i}^2 \ln \mathcal{L}(\params_\ml) \, \partial_{\param_j}^2 \ln \mathcal{L}(\params_\ml) - \left(\partial_{\param_i}\partial_{\param_j} \ln \mathcal{L}(\params_\ml)\right)^2
\end{equation}
When ignoring the covariances, the cross-terms vanish and the naive error estimate for $\param_i$ reduces to
\begin{equation}
\tilde{\sigma}_{\param_i}^2(\params_\ml) = -\left(\partial^2_{\param_i} \ln \mathcal{L}(\params_\ml)\right)^{-1}
\end{equation}
which is identical to our results from \S\ref{sec:preliminaries}. Properly including the covariance instead gives
\begin{equation}
\sigma_{\param_i}^2(\params_\ml) = \frac{-\partial^2_{\param_j} \ln \mathcal{L}(\params_\ml)}{\partial_{\param_i}^2 \ln \mathcal{L}(\params_\ml) \, \partial_{\param_j}^2 \ln \mathcal{L}(\params_\ml) - \left(\partial_{\param_i}\partial_{\param_j} \ln \mathcal{L}(\params_\ml)\right)^2}
\end{equation}
}

\edit1{
The ratio of the naive estimate to the true estimate of the variance is
\begin{equation}
\frac{\tilde{\sigma}_{\param_i}^2(\params_\ml)}{\sigma_{\param_i}^2(\params_\ml)} = \frac{\partial_{\param_i}^2 \ln \mathcal{L}(\params_\ml) \, \partial_{\param_j}^2 \ln \mathcal{L}(\params_\ml) - \left(\partial_{\param_i}\partial_{\param_j} \ln \mathcal{L}(\params_\ml)\right)^2}{\partial_{\param_i}^2 \ln \mathcal{L}(\params_\ml) \, \partial_{\param_j}^2 \ln \mathcal{L}(\params_\ml)} 
= 1 - \frac{\left(\partial_{\param_i}\partial_{\param_j} \ln \mathcal{L}(\params_\ml)\right)^2}{\partial_{\param_i}^2 \ln \mathcal{L}(\params_\ml) \, \partial_{\param_j}^2 \ln \mathcal{L}(\params_\ml)}
\end{equation}
This gives a fractional bias of
\begin{equation}
\boxed{
\frac{\delta_{\tilde{\sigma}_{\param_i}^2}}{\sigma_{\param_i}^2}(\params_\ml) \equiv \frac{\tilde{\sigma}_{\param_i}^2(\params_\ml)}{\sigma_{\param_i}^2(\params_\ml)} - 1 =  -\frac{\left(\partial_{\param_i}\partial_{\param_j} \ln \mathcal{L}(\params_\ml)\right)^2}{\partial_{\param_i}^2 \ln \mathcal{L}(\params_\ml) \, \partial_{\param_j}^2 \ln \mathcal{L}(\params_\ml)} 
}
\end{equation}
This is defined to be negative to indicate that we're dealing with underestimates rather than overestimates.\footnote{This is a consequence of the fact that $\left(\partial_{\param_i}\partial_{\param_j} \ln \mathcal{L}\right)^2 \leq (\partial_{\param_i}^2 \ln \mathcal{L})(\partial_{\param_j}^2 \ln \mathcal{L})$, which arises from the Cauchy-Schwarz inequality.}
}

\edit1{
This result -- where the error underestimates are the ratio of the product of the ``interaction'' terms divided by the ``naive'' terms -- provides a quick and intuitive way to estimate how covariances among parameters impact our marginalized error estimates.\footnote{Unfortunately, this intuition does not generalize to larger matrices.} In the case where there is no covariance among the parameters, $\delta_{\tilde{\sigma}_{f}^2}/\sigma_{f}^2 = 0$ and $\tilde{\sigma}^2_f = \sigma^2_f$. In the case where our parameters become perfectly degenerate, we instead get $\delta_{\tilde{\sigma}_{f}^2}/\sigma_{f}^2 \rightarrow 1$ and $\sigma_f^2/\tilde{\sigma}_f^2 \rightarrow \infty$.
}

\subsection{Background Covariance}
\label{subsec:fisher_fb}

\edit1{
As an example, consider that we fix the position $(x,y)$ of the source and are jointly estimating the flux density $f$ and sky background $b$. The mixed $f$ and $b$ derivative is
\begin{align}
\partial_f\partial_b\ln\mathcal{L}(x,y) = -\frac{1}{\sigma^2}\sum_i p_i(x,y) = -\frac{1}{\sigma^2}
\end{align}
which is insensitive to the value of $f$ and $b$. This then gives
\begin{equation}
\fisher_{\params}(x,y) = -\frac{1}{\sigma^2}
\begin{bmatrix}
\neff(x,y) & 1 \\
1 & A \\
\end{bmatrix}
\end{equation}
where again $\neff(x,y) = \sum_i p_i^2(x,y)$ and $A = \xdim\ydim$. This in turn gives
\begin{equation}
\cov_{\params}(x,y) = \frac{\sigma^2}{A \neff(x,y) -1}
\begin{bmatrix}
A & -1 \\
-1 & \neff(x,y) \\
\end{bmatrix}
\end{equation}
This means the naive errors $\tilde{\sigma}_f^2$ we previously derived, which ignored the covariance between the flux and the background, are smaller than the actual uncertainties by a factor
\begin{equation}
\boxed{\frac{\delta_{\tilde{\sigma}_{f}^2}}{\sigma_{f}^2}(x,y) = -\frac{\neff(x,y)}{A}}
\end{equation}
}

\subsection{Position Covariance}
\label{subsec:fisher_fx}

\edit1{
The mixed partial derivative of the log-likelihood with respect to $f$ and one coordinate of the position, say, $x$, is
\begin{align}
\partial_f\partial_x\ln\mathcal{L}(x,y,f,b) = \frac{1}{\sigma^2} \sum_i \left(\hat{f}_i - b - f p_i(x,y) - f p_i(x,y)\right) \partial_x p_i(x,y) 
\end{align}
In the case where $\psf_{x,y}$ is approximately symmetric (even) in $x$, the derivative $\partial_x \psf_{x,y}$ will be approximately antisymmetric (odd) in $x$. We would then expect
\begin{equation}
\sum_i p_i(x,y) \partial_x p_i(x,y) \approx \iint p(x,y) \partial_x p(x,y)\;dx\;dy = 0
\end{equation}
assuming that (1) the PSF is oversampled so that our sum over pixels is a reasonable approximation to the integral and (2) the impact of sub-pixel shifts is small. If our PSF is undersampled so that it occupies only a few pixels, then this term may be significantly non-zero. Note that the behavior under any general covariance matrix is not guaranteed to be small even if the PSF is oversampled, although in most practical applications where only nearby pixels are correlated with each other this behavior still tends to hold.
}

\edit1{
If our model $(x_\ml, y_\ml, f_\ml)$ is close to the truth, we expect the residuals $\hat{\flux} - \bflux - f\psf(x,y)$ to be roughly distributed following a multivariate Normal, which we will write as
\begin{equation}
\mathbf{X} \sim \cov^{1/2} \mathbf{Z} \sim \mathcal{N}(\zeros, \cov)
\end{equation}
where $\normvec$ is an $mn \times 1$ iid normal \textbf{random vector} and $\cov^{1/2}$ is the symmetric square root of the covariance matrix $\cov$. This implies that the residual contribution to our flux density-position covariance should be roughly distributed as
\begin{equation}
\left(\cov^{1/2} \normvec\right)^\T \cov^{-1} \partial_x \psf_{x,y} = \normvec^\T \cov^{-1/2} \partial_x \psf_{x,y}
\end{equation}
Since the expectation value $\mathbb{E}[\normvec^\T \mathbf{Y}] = \mathbb{E}[\normvec]^\T \mathbf{Y} = \zeros$ for a normal random vector given any fixed matrix $\mathbf{Y}$, our FIM reduces to
\begin{equation}
\fisher_{\params}(x_\ml,y_\ml,f_\ml) \approx -\frac{1}{\sigma^2}
\begin{bmatrix}
\sum_i p_i^2(x_\ml,y_\ml) & 0 \\
0 & f_\ml^2 \sum_i (\partial_x p_i(x_\ml,y_\ml))^2 \\
\end{bmatrix}
\end{equation}
and
\begin{equation}
\boxed{\frac{\delta_{\tilde{\sigma}_{f}^2}}{\sigma_{f}^2}(x_\ml, y_\ml, f_\ml) \approx 0}
\end{equation}
This implies that while modeling the position leads to a bias in the mean value of $f_\ml(x_\ml, y_\ml)$, it \textit{does not} impact the associated error estimates.
}

\section{Errors with General Background Models}
\label{ap:back_gen}

In \S\ref{subap:gen_res_back}, we showed that the maximum-likelihood (ML) solution for a background model $\bflux$ across all $\xdim\ydim$ pixels is
\begin{equation}
\bflux_\ml(x,y,f) = \hat{\flux} - f\psf_{x,y}
\end{equation}
which had an error estimate of
\begin{equation}
\tilde{\cov}_\bflux = \cov
\end{equation}
This result is singularly uninformative, because it implies that the ``best'' background model is exactly equal to the model residuals across the entire image.

In most cases, we often seek to parameterize the background as a smooth function $\bflux_{\bparams} \equiv \bflux(\bparams)$ of $\bdim$ nuisance parameters $\bparams$ across the footprint. This gives us
\begin{equation}
\partial_{\beta_i}\ln\mathcal{L}(x,y,f,\bparams_\ml) = \partial_{\beta_i} \bflux_{\bparams_\ml}^\T \cov^{-1} (\hat{\flux} - f\psf_{x,y} - \bflux_{\bparams_\ml}) = 0
\end{equation}
which we can use to solve for $\bparams_\ml$. Following Appendix \ref{ap:error}, the error estimates can be derived by inverting the Fisher Information Matrix (FIM) whose elements are
\begin{equation}
(\fisher_{\bparams})_{ij}(x,y,f,\bparams_\ml) = \partial_{\beta_i} \bflux_{\bparams_\ml}^\T \cov^{-1} \partial_{\beta_j} \bflux_{\bparams_\ml} - \partial_{\beta_i}\partial_{\beta_j} \bflux_{\bparams_\ml}^\T \cov^{-1} (\hat{\flux} - f\psf_{x,y} - \bflux_{\bparams_\ml})
\end{equation}
As before, if we assume that our overall residuals $\hat{\flux} - f\psf_{x,y} - \bflux_{\bparams_\ml}$ are small and that our background model varies sufficiently slowly with respect to $\bparams$, we can approximate this as
\begin{equation}
\boxed{\fisher_{\bparams}(\bparams_\ml) \approx \partial_{\bparams} \bflux_{\bparams_\ml}^\T \cov^{-1} \partial_{\bparams} \bflux_{\bparams_\ml}}
\end{equation}
where $\partial_{\bparams} \bflux_{\bparams_\ml}$ is the $\xdim\ydim \times \bdim$ Jacobian. This implies that we can use the Jacobian to linearly project from the $\xdim\ydim$-dimensional ``data space'' into the $\bdim$-dimensional parameter space.

\section{Maximum-Likelihood Biases using Cochran's Theorem}
\label{ap:bias}

As discussed in \S\ref{subsec:bias_gen}, given the true values $(x^*,y^*,f^*,\bflux^*)$ of the position, flux density, and background, respectively, we can rewrite the likelihood of the noisy data as
\begin{equation}
\ln \mathcal{L}(x^*,y^*,f^*,\bflux^*) = -\frac{1}{2} \ln(\det(2\pi \cov)) -\frac{1}{2} \sum_i Z_i^2
\end{equation}
where $Z_1, \dots, Z_i \sim \mathcal{N}(0, 1)$ are again iid normal random variables and
\begin{equation}
\sum_{i=1}^{\xdim\ydim} Z_i^2 \sim \chi^2_{A}
\end{equation}
follows a chi-square distribution with $A=\xdim\ydim$ degrees of freedom. Assuming that the data are normally distributed and our ML parameters are also approximately Normally distributed, we can apply Cochran's theorem to get
\begin{equation}
(\hat{\flux} - \flux_{\params_\ml})^\T \cov^{-1} (\hat{\flux} - \flux_{\params_\ml}) \sim \chi^2_{A-p}
\end{equation}

Assuming the background is known (see \S\ref{subsubsec:back_coup}), we note that we can relate the distribution of the sum of error normalized residuals around $f_\ml \equiv f_\ml(x_\ml, y_\ml)$ in the decoupled-background case to those around $ f_\ml^* \equiv f_\ml(x^*, y^*)$ for a constant background model $b$ via
\[
(\hat{\flux} - f_\ml\psf_{x_\ml,y_\ml})^\T \cov^{-1} (\hat{\flux} - f_\ml\psf_{x_\ml,y_\ml}) + X^2_2 \sim (\hat{\flux} - f_\ml^*\psf_{x^*,y^*})^\T \cov^{-1} (\hat{\flux} - f_\ml^*\psf_{x^*,y^*})
\]
where $X^2_2 \sim \chi^2_2$ incorporates the noise ``absorbed'' by $(x_\ml, y_\ml)$ and we have exploited the fact that $X^2_2 + X^2_{\xdim\ydim-2} \sim \chi^2_{\xdim\ydim}$. Exploiting the fact that
\begin{equation}
\ln \mathcal{L}(x,y,f_\ml) = \frac{f_\ml^2(x,y)}{2\tilde{\sigma}_f^2(x,y)}
\end{equation}
then allows us to rewrite the above result as
\begin{equation}
\hat{\flux}^\T \cov^{-1}\hat{\flux} - 2 f_\ml \hat{\flux}^\T \cov^{-1} \psf_{x_\ml, y_\ml} + \frac{f_\ml^2}{\tilde{\sigma}_{f_\ml}^2} + X^2_2 \sim \hat{\flux}^\T \cov^{-1} \hat{\flux} - 2 f_\ml^* \hat{\flux}^\T \cov^{-1} \psf_{x^*, y^*} + \frac{f_\ml^{*,2}}{\tilde{\sigma}_{f_\ml^*}^2}
\end{equation}
Although the true position is not known, in the interest of deriving the impact on the flux we can take the approximation that $(x^*, y^*) \approx (x_\ml, y_\ml)$, etc. for all terms that don't explicitly involve the flux density $f$. This leaves us with an equation of the form
\begin{equation}
c_1 f_\ml^2 - 2 c_2 f_\ml + X^2_2 \sim c_1 (f_\ml^*)^2 - 2 c_2 f_\ml^*
\end{equation}
where $c_1 = \tilde{\sigma}_{f_\ml}^{-2}$ and $c_2 = \hat{\flux}^\T \cov^{-1} \psf_{x_\ml, y_\ml}$ are roughly constant. This has a positive solution at
\begin{equation}
f_\ml^* \sim \frac{c_2 + \sqrt{c_1^2f_\ml^2 - c_1X^2_2 - 2c_1c_2f_\ml + c_2^2}}{c_1} = \frac{c_2}{c_1} + f_\ml\sqrt{1 - \frac{2c_2}{c_1f_\ml} - \frac{X^2_2}{c_1f_\ml^2} + \frac{c_2^2}{c_1^2f_\ml^2}}
\end{equation}
Since $c_1$ and $c_2$ are known for a given $(x_\ml, y_\ml)$ and the distribution of $X^2_2 \sim \chi^2_2$ is known exactly, this gives an expression for the distribution of the unbiased ML estimator $f_\ml^*$. In general, assuming that the residuals are sufficiently small such that $c_2 \approx 0$, this reduces to
\begin{equation}
f_\ml^* \sim f_\ml\sqrt{1 - X^2_2\frac{\tilde{\sigma}_{f_\ml}^2}{f_\ml^2}}
\end{equation}
This can be immediately generalized to a model with $p$ model parameters $\params$ (excluding the background $b$) to get
\begin{equation}
f_\ml^* \sim f_\ml\sqrt{1 - X^2_{p-1}\frac{\tilde{\sigma}_{f_\ml}^2}{f_\ml^2}}
\end{equation}

We can write this in a slightly more intuitive form by Taylor expanding around small $\tilde{\sigma}_{f_\ml}/f_\ml$ to get
\begin{equation}
f_\ml^* \approx f_\ml \left[ 1 - \frac{X^2_{p-1}}{2}\frac{\tilde{\sigma}^2_{f_\ml}}{f_\ml^2} \right]
\end{equation}
This gives a fractional bias of
\begin{equation}
\boxed{
1 - \frac{\mathbb{E}[f_\ml^*]}{f_\ml} \equiv \frac{\delta_{f_\ml}}{f_\ml} \approx \frac{p-1}{2}\frac{\tilde{\sigma}^2_{f_\ml}}{f_\ml^2} \,,~ \frac{\mathbb{V}[f_\ml^*]}{\tilde{\sigma}_{f_\ml}^2} \approx \frac{p^2-4p+7}{4}\frac{\sigma^2_{f_\ml}}{f_\ml^2}
}
\end{equation}

At lower SNR it is necessary to include the second-order term $\frac{X_{p-1}^4}{8}\frac{\sigma^4_{f_\ml}}{f_\ml^4}$ from the Taylor expansion to properly model behavior. Including this term gives a modified fractional bias of
\begin{equation}
\frac{\delta_{f_\ml}}{f_\ml} \approx \frac{p-1}{2}\frac{\tilde{\sigma}^2_{f_\ml}}{f_\ml^2} + \frac{p^2-1}{8}\frac{\tilde{\sigma}^4_{f_\ml}}{f_\ml^4}
\end{equation}
For a typical point source model with $p=3$ parameters $(x, y, f)$, this becomes
\begin{equation}
\frac{\delta_{f_\ml}}{f_\ml} \approx \frac{\tilde{\sigma}^2_{f_\ml}}{f_\ml^2} + \frac{\tilde{\sigma}^4_{f_\ml}}{f_\ml^4}
\end{equation}

\section{Maximum-Likelihood Biases using Bias Tensors}
\label{ap:bias_tensor}

ML estimators have a bias $\delta$ which tends to zero as the signal-to-noise ratio (SNR) increases. \cite{10.2307/2984505} found that the leading-order bias term for any parameter $s$ can be found with
\begin{equation}
\bias_s(\params_\ml) = \sum_{r,t,u} (\fisher^{-1}(\params_\ml))_{rs} \, (\fisher^{-1}(\params_\ml))_{tu} \, (\btensor(\params_\ml))_{rtu}
\end{equation}
where 
\begin{equation}
(\btensor(\params_\ml))_{rtu} \equiv \mathbb{E}_{\mathbf{D}}\left[\frac{1}{2} \partial_r\partial_t\partial_u\ln \mathcal{L}(\params_\ml) + (\partial_t \ln \mathcal{L}(\params_\ml)) (\partial_r\partial_u \ln \mathcal{L}(\params_\ml)) \,\middle|\, \params_\ml \right]
\end{equation}
is the bias tensor and $\mathbb{E}_\mathbf{D}[\cdot|\params_\ml]$ is the expectation value with respect to the data $\mathbf{D}$ for $\params_\ml$ fixed.

With the background $b=b^*$ fixed, the non-zero terms in the flux density bias $\bias_f$ are
\begin{equation}
\label{eqn:biastensor_backfixed}
\bias_f(\params_\ml) = \tilde{\sigma}_{f_\ml}^2 \sum_{i\in \{x,y\}} \sigma_x^2 \, (\btensor(\params_\ml))_{fii}
\end{equation}
since the off-diagonal elements of the FIM with respect to position $(x,y)$ are zero (\S\ref{subsec:fisher_fx}) and we have substituted in for $(\fisher^{-1}(\params_\ml))_{ff} = \tilde{\sigma}_{f_\ml}^2$ and $(\fisher^{-1}(\params_\ml))_{xx} = (\fisher^{-1}(\params_\ml))_{yy} = \sigma_x^2$.
Under similar assumptions as \S\ref{subsec:bias_gen}, is it straightforward to show that
\begin{equation}
\label{eqn:biastensor_elements}
(\btensor(\params_\ml))_{fxx} = (\btensor(\params_\ml))_{fyy} = \frac{1}{2 \sigma_x^2 f}
\end{equation}
and thus:
\begin{equation}
\bias_{f_\ml} = \frac{\tilde{\sigma}_{f_\ml}^2}{f_\ml}.
\end{equation}
Substituting in for the definition of $\bias_f$
allows us to rewrite this as
\begin{equation}
\mathbb{E}[f_\ml^*] = f_\ml \left(1-\frac{\tilde{\sigma}^2_{f_\ml}}{f_\ml^2}\right)
\end{equation}
which reproduces equation \eqref{eqn:bias_ideal}.

With the background $b$ free, the flux bias has the same terms as Equation \ref{eqn:biastensor_backfixed} except that $(\fisher^{-1}(\params_\ml))_{ff} = \sigma_{f_\ml}^2 \geq \tilde{\sigma}_{f_\ml}^2$ due to the covariance between the flux and background. Solving and rearranging as above then gives
\begin{equation}
\boxed{\mathbb{E}[f_\ml^*] = f_\ml \left(1-\frac{\sigma^2_{f_\ml}}{f_\ml^2}\right)}
\end{equation}
which reproduces equation \eqref{eqn:bias_point_source}.

\section{Unbiasedness of Marginalized Mean Flux}
\label{ap:flux_mean}

While the ML solution is the mode of the likelihood distribution, the \textit{mean} flux density $f_{\rm mean}$, marginalizing over position $(x,y)$, need not be the ML flux density $f_\ml$. Here, we show that the posterior mean flux is unbiased to order $SNR^{-2}$ in a specific case (flat priors and Gaussian PSF) for illustrative purposes.

As discussed in \S\ref{sec:preliminaries}, the maximum likelihood flux $f_\ml(x,y)$ varies with position and is maximized at $(x_\ml,y_\ml)$ assuming the background $b$ is known. While first derivative of $f_\ml \equiv f_\ml(x_\ml, y_\ml)$ with position is zero (by definition), the second derivative is
\begin{equation}
\partial_x^2 f_\ml \approx \frac{1}{\sum_j p_j^2} \sum_i \partial^2_x p_i \, \hat{f}_i
\end{equation}
under the same oversampled/smoothness assumptions discussed in \S\ref{subsec:ml_pos}. For a circular Gaussian PSF with standard deviation of $s$ pixels, the second derivative of the PSF is
\begin{equation}
\partial^2_x p_i = \frac{\Delta x_i^2-s^2}{s^4} p_i
\end{equation}
where $\Delta x_i$ is the difference in the x-coordinate of the center of pixel $i$ and the source. Taking
\begin{equation}
\frac{\sum_i \Delta x_i^2 p_i^2}{\sum_j p_j^2} \approx \frac{s^2}{2}
\end{equation}
and approximating the counts around the source as $\hat{f}_i \approx f_\ml p_i$ then gives
\begin{equation}
\partial_x^ 2f_\ml \approx -\frac{f_\ml}{2 s^2}.
\end{equation}
using the definition of $f_\ml$ from equation \eqref{eqn:ml_flux}.

\textit{Near} the ML position, the ML flux density is approximately
\begin{equation}
\label{eqn:best_fit_flux_pos}
f_\ml(x,y) \approx f_\ml \, \left[1-\frac{(x-x_\ml)^2+(y-y_\ml)^2}{4 s^2}\right]
\end{equation}
The likelihood can then be approximated (up to a constant factor $Z$) as
\begin{equation}
\mathcal{L}(x,y,f) = Z \exp\left(-\frac{(x-x_\ml)^2+(y-y_\ml)^2}{2 \tilde{\sigma}_x^2(f_\ml)}-\frac{(f - f_\ml(x,y))^2}{2\tilde{\sigma}_f^2(f_\ml)}\right)
\end{equation}
where we have taken $\tilde{\sigma}_x^2 = \tilde{\sigma}_y^2$ since our PSF is circular.
The mean flux density $f_{\rm mean}$ after marginalizing over position is defined as
\begin{equation}
f_{\rm mean} \equiv \frac{\iiint \mathcal{L}(x,y,f)\,f\;df\;dx\;dy}{\iiint \mathcal{L}(x,y,f)\;df\;dx\;dy}
\end{equation}
Integrating over the flux density $f$ yields
\begin{equation}
f_{\rm mean} \approx \frac{\iint f_\ml(x,y) \exp\left(-\frac{(x-x_\ml)^2+(y-y_\ml)^2}{2 \sigma_x^2(f_\ml)}\right) \;dx\;dy}{\iint \exp\left(-\frac{(x-x_\ml)^2+(y-y_\ml)^2}{2 \sigma_x^2(f_\ml)}\right) \;dx\;dy}
\end{equation}
and then using equation \eqref{eqn:best_fit_flux_pos} for $f_\ml(x,y)$ gives
\begin{equation}
f_{\rm mean} \approx f_\ml \left(1-\frac{\sigma_x^2(f_\ml)}{2 s^2}\right)
\end{equation}
Substituting in $\sigma_x^2(f) = 2 \sigma_f^2 s^2 / f^2$ for our circular Gaussian PSF then gives
\begin{equation}
\boxed{f_{\rm mean} \approx f_\ml \left(1 - \frac{\sigma_{f_\ml}^2}{f_\ml^2}\right) = \mathbb{E}[f^*_\ml]}
\end{equation}
where $\sigma_{f_\ml} \equiv \sigma_f(f_\ml)$ and $f^*_\ml = f_\ml(x^*, y^*)$ are defined as in \S\ref{sec:bias}.

Since we showed \S\ref{sec:bias} that the ML estimator at the true position $f_\ml^*$ is unbiased, this explicitly demonstrates that the mean flux density, marginalized over position, is also unbiased.

\section{Second-Order Expansion in Flux Density}
\label{ap:flux_taylor}

As discussed in Section \S\ref{subsec:phot_single}, fixing the position of an object to the best-fit from a detection band $D$ leads to an underestimate in the flux density in all other bands. If the detected position is distributed as a 2-D Gaussian of width $\sigma_x$ about the true position, the average best-fit flux is given by Equation \ref{eqn:detect_underest_gaussian}: 

\begin{equation}
\mathbb{E}[f_{j,\ml}] = f_j^* \times \frac{2 s^2}{2 s^2 + \sigma_{D,x}^2}
\end{equation}

At low signal to noise, the position error obtained by inverting the second derivative with respect to position (Equation \ref{eqn:position_error}) needs to be corrected by higher-order terms. The maximum-likelihood is found by setting the partial derivatives of the log-likelihood to zero. We Taylor expand the partial derivatives to leading order about the true parameters $(x^*, y^*, f^*)$:
\begin{equation}
\label{eqn:max_like_order_i}
0 = \partial_\alpha\ln\mathcal{L}(x_\ml,y_\ml,f_\ml) \approx \partial_\alpha\ln\mathcal{L}(x^*,y^*,f^*) +  \sum_\beta (\beta_\ml - \beta^*) \partial^2_{\alpha\beta}\ln\mathcal{L}(x^*,y^*,f^*)
\end{equation}
with $\alpha,\beta\in \{x,y,f\}$. Evaluating the partial derivatives at the true parameters yields
\begin{equation}
\label{eqn:first_derivatives}
\partial_\alpha\ln\mathcal{L}(x^*,y^*,f^*) = \frac{1}{\sigma^2} \sum_i \sigma Z_i \partial_\alpha(f p_i(x,y))|_{\params=\params^*}
\end{equation}
\begin{equation}
\label{eqn:second_derivatives}
\partial_{\alpha\beta}\ln\mathcal{L}(x^*,y^*,f^*) =
\frac{1}{\sigma^2} \sum_i \sigma Z_i \partial_{\alpha\beta}(f p_i(x,y))|_{\params=\params^*} - \partial_{\alpha}(f p_i(x,y))|_{\params=\params^*}\partial_{\beta}(f p_i(x,y))|_{\params=\params^*}
\end{equation}
with $Z_i = (\hat{f}_i - f^* p_i(x^*,y^*))$ being independent, normally distributed variables with mean zero and unit variance. The sum over pixels of the second term of Equation \ref{eqn:second_derivatives} can be approximated in the oversampled limit with integrals over the PSF and its derivatives. Using Equations \ref{eqn:first_derivatives} and \ref{eqn:second_derivatives}, Equation \ref{eqn:max_like_order_i} can be written as a matrix equation:
\begin{equation}
0 =
\sigma
\begin{bmatrix}
\zeta \\
f^* \zeta_x \\
f^* \zeta_y
\end{bmatrix}
+
\left(
\sigma
\begin{bmatrix}
0 & \zeta_x & \zeta_y \\
\zeta_x & f^* \zeta_{xx} & f^* \zeta_{xy} \\
\zeta_y & f^* \zeta_{xy} & f^* \zeta_{yy}
\end{bmatrix}
-
\begin{bmatrix}
\frac{1}{4\pi s^2} & 0 & 0 \\
0 & \frac{f^*}{8\pi s^4} & 0 \\
0 & 0 & \frac{f^*}{8\pi s^4}
\end{bmatrix}
\right)
\begin{bmatrix}
f_\ml - f^* \\
x_\ml - x^* \\
y_\ml - y^*
\end{bmatrix}
\end{equation}
with $\zeta = \sum_i Z_i p_i(x^*,y^*)$, $\zeta_\alpha = \sum_i Z_i \partial_\alpha p_i(x^*,y^*)$, and $\zeta_{\alpha\beta} = \sum_i Z_i \partial_{\alpha\beta} p_i(x^*,y^*)$.
Solving this matrix equation for $x_\ml - x^*$ and expanding to second order in $\sigma$ yields:
\begin{equation}
\label{eqn:xml_order_i}
x_\ml - x^* = \frac{8\pi s^4 \sigma \zeta_x}{f^*} + \frac{32 \pi^2 s^6 \sigma^2 (\zeta \zeta_x + 2 s^2[\zeta_x \zeta_{xx} + \zeta_y \zeta_{xy}])}{f^{*2}}.
\end{equation}
The sums $\zeta, \zeta_\alpha, \zeta_{\alpha\beta}$ are random variables with mean zero just as $Z_i$ are. It can be shown that the expectation value of $x_\ml - x^*$ is zero (ie. the position is unbiased) by evaluating expectation values like
\begin{equation}
\langle\zeta\zeta_x\rangle = \left\langle \sum_{ij} Z_i p_i(x^*,y^*) Z_j \partial_x p_i(x^*,y^*) \right\rangle = \sum_i \langle Z_i^2 \rangle p_i(x^*,y^*) \partial_x p_i(x^*, y^*) \approx \iint p(x,y) \partial_x p(x,y) dx\;dy= 0
\end{equation}
with the second equality following from the independence of the $Z_i$ and the third equality holding in the oversampled limit. Similarly,
\begin{equation}
\langle\zeta_x\zeta_{xx}\rangle = \langle\zeta_y\zeta_{xy}\rangle = 0
\end{equation}
Since $x_\ml - x^*$ has an expectation value of zero, the variance in position can be found by squaring Equation \ref{eqn:xml_order_i} and taking the expectation value. To evaluate the variance in position, these expectation values are needed:
\begin{align}
\langle\zeta^2\rangle = \frac{1}{4\pi s^2} \\
\langle\zeta_x^2\rangle = \langle\zeta_y^2\rangle = \frac{1}{8\pi s^4} \\
\langle\zeta_{xx}^2\rangle = \frac{3}{16\pi s^6} \\
\langle\zeta_{xy}^2\rangle = \frac{1}{16\pi s^6} \\
\langle\zeta \zeta_{xx}\rangle = -\frac{1}{8\pi s^4} \\
\langle\zeta \zeta_x^2\rangle = \langle\zeta\rangle \langle\zeta_x^2\rangle = 0 \\
\langle\zeta_x^2 \zeta_{xx} \rangle = \langle\zeta_x^2\rangle \langle\zeta_{xx}\rangle = 0 \\
\langle\zeta_x \zeta_y \zeta_{xy}\rangle = \langle\zeta_x\rangle \langle\zeta_y\rangle \langle\zeta_{xy}\rangle = 0 \\
\langle\zeta^2 \zeta_x^2\rangle = \langle\zeta^2\rangle \langle\zeta_x^2\rangle \\
\langle\zeta \zeta_x^2 \zeta_{xx}\rangle = \langle\zeta \zeta_{xx}\rangle \langle\zeta_x^2\rangle \\
\langle\zeta \zeta_x \zeta_y \zeta_{xy}\rangle = \langle\zeta \rangle \langle\zeta_x\rangle \langle\zeta_y\rangle \langle\zeta_{xy}\rangle = 0\\
\langle\zeta_x^2 \zeta_{xx}^2\rangle = \langle\zeta^2\rangle \langle\zeta_{xx}^2\rangle \\
\langle\zeta_x \zeta_y \zeta_{xx} \zeta_{xy}\rangle = \langle\zeta_x\rangle \langle\zeta_y\rangle \langle\zeta_{xx}\rangle \langle\zeta_{xy}\rangle = 0 \\
\langle\zeta_y^2 \zeta_{xy}^2\rangle = \langle\zeta_y^2\rangle \langle\zeta_{xy}^2\rangle
\end{align}
which yield:
\begin{equation}
\langle(x_\ml-x^*)^2\rangle = \frac{8\pi s^4 \sigma^2}{f^{*2}}\left(1 + \frac{28 \pi s^2 \sigma^2}{f^{*2}}\right) = \frac{2 s^2 \tilde{\sigma}_f^2}{f^{*2}} \left(1 + \frac{7 \tilde{\sigma}_f^2}{f^{*2}}\right)
\end{equation}
Approximating the position errors as being distributed as a 2-D Gaussian with this variance and using Equation \ref{eqn:detect_underest_gaussian} gives:
\begin{equation}
\mathbb{E}[f_{j,\ml}] = f_j^* \times \frac{2s^2}{2 s^2 + \frac{2 s_D^2 \tilde{\sigma}_{f_D}^2}{f_D^{*2}} \left(1 + \frac{7 \tilde{\sigma}_{f_D}^2}{f_D^{*2}}\right)} \approx \boxed{ f_j^* \left(1 - \frac{s_D^2}{s^2} \frac{\tilde{\sigma}_{f_D}^2}{f_D^{*2}} - \left[\frac{7 s_D^2}{s^2} - \frac{s_D^4}{s^4}\right] \frac{\tilde{\sigma}_{f_D}^4}{f_D^{*4}}\right) }
\end{equation}

\section{Gaussian Galaxies}
\label{ap:gaussgal}

\edit1{
To illustrate the impact extended sources such as galaxies can have on the ML flux bias (by having more free parameters, additional parameter covariances, and enlarged effective PSFs), we will derive results explicitly below for a circular Gaussian PSF with variance $s^2$ and a 2-D Gaussian galaxy model with semi-major/semi-minor axes $s_1$/$s_2$ and position angle $\phi$. Without loss of generality, we will take the true $\phi^* = 0$ to simplify our calculations since it doesn't affect the size of the galaxy. These results build on those from \cite{1997PASP..109..166C} and \cite{2012MNRAS.425.1951R}.
}

\edit1{
For a point source, the effective PSF area is 
\begin{equation}
\neff = \sum_i p_i^2(x,y) \rightarrow 4\pi s^2
\end{equation} in the limit where the footprint is sufficiently large and the PSF is oversampled. For a Gaussian galaxy convolved with our PSF, however, this increases to 
\begin{equation}
\neff \rightarrow 4\pi \sqrt{(s_1^2+s^2) (s_2^2+s^2)} \equiv 4\pi a_1 a_2
\end{equation} 
where $a_1$ and $a_2$ are now the PSF-convolved semi-major and semi-minor axes, respectively. Assuming that $a_1=a_1^*$, $a_2=a_2^*$, and $\phi=\phi^*$ are \textit{known and fixed} to their true values, this would increase the underlying bias by
\begin{equation}
\frac{\delta_{f_\ml}}{f_\ml} \approx \frac{2}{2} \frac{{\sigma}^2_{f_\ml}}{f_\ml^2} = \frac{1}{1-\frac{\neff}{A}}\frac{\tilde{\sigma}^2_{f_\ml}}{f_\ml^2} = \frac{\neff}{1 - \frac{\neff}{A}}\frac{\sigma^2}{f_\ml^2}
\end{equation}
where $\delta_f/f$ is again the fractional bias and $\sigma^2$ is again the variance of the normal iid noise in the footprint. At fixed SNR, this then implies
\begin{equation}
\frac{\delta_{f_\ml}}{f_\ml}(a_1^*, a_2^*, \phi^*, A) = \frac{\neff(a_1^*, a_2^*)}{1 - \frac{\neff(a_1^*, a_2^*)}{A}}\frac{\sigma^2}{f_\ml^2} = \frac{4\pi a_1^*a_2^*}{1 - \frac{4\pi a_1^*a_2^*}{A}}\frac{\sigma^2}{f_\ml^2}
\end{equation}
}

\edit1{
While the case above is instructive, it is not representative of the general case where $a_1$, $a_2$, and $\phi$ are left free when searching for a ML solution. This leads to additional covariances among the parameters that cause the uncertainties to increase. To find these covariances, we need second partial derivatives of the log-likelihood
\begin{equation}
\partial_{\alpha\beta}\ln\mathcal{L} =
\frac{1}{\sigma^2} \sum_i (\hat{f}_i - f p_i(x,y) - b) \partial_{\alpha\beta}(f p_i(x,y) + b) - \partial_{\alpha}(f p_i(x,y) + b) \partial_{\beta}(f p_i(x,y) + b)
\end{equation}
with $p_i$ being the galaxy model convolved with the PSF and $\alpha,\beta$ indexing the parameters of the galaxy. To fill in the FIM, we take the expectation value of the second derivatives with the parameters fixed to their true values. The residuals about the true parameters $\hat{f}_i - f p_i$ are zero in expectation, so
\begin{equation}
\fisher_{\alpha\beta} = \frac{1}{\sigma^2} \sum_i \partial_{\alpha}(f p_i(x,y) + b) \partial_{\beta}(f p_i(x,y) + b) \approx \iint \partial_{\alpha}(f p(x,y) + b) \partial_{\beta}(f p(x,y) + b) \;dx\;dy 
\end{equation}
in the well-sampled limit, with
\begin{equation}
p(x,y) = \frac{1}{2\pi a_1 a_2} \exp\left(-\frac{(x \cos \varphi + y \sin \varphi)^2}{2a_1^2}-\frac{(-x\sin\varphi+y\cos\varphi)^2}{2a_2^2}\right)
\end{equation}
Many of the off-diagonal elements are zero due to the symmetries of $p(x,y)$, so the FIM is diagonal except for the entries involving $f$, $b$, $a_1$, and $a_2$. The FIM for these four parameters is
\begin{equation}
\fisher = 
\frac{1}{\sigma^2}
\begin{bmatrix}
    \frac{1}{4\pi a_1 a_2} & 1 & -\frac{f}{8\pi a_1^2 a_2} & -\frac{f}{8\pi a_1 a_2^2} \\
    1 & A & 0 & 0 \\
    -\frac{f}{8\pi a_1^2 a_2} & 0 & \frac{3 f^2}{16\pi a_1^3 a_2} & \frac{f^2}{16\pi a_1^2 a_2^2} \\
    -\frac{f}{8\pi a_1 a_2^2} & 0 & \frac{f^2}{16\pi a_1^2 a_2^2} & \frac{3 f^2}{16\pi a_1 a_2^3}
\end{bmatrix}
\end{equation}
Note that $\fisher_{ff} = (\neff \sigma^2)^{-1}$ for a galaxy and so without accounting for the covariances, $\tilde{\sigma}_f^2 = \neff \sigma^2$. Taking the covariances into account by inverting the FIM yields:
\begin{equation}
(\fisher^{-1})_{ff} = \frac{8 \pi a_1 a_2 \sigma^2}{1-8 \pi a_1 a_2/A} = \frac{2 \neff \sigma^2}{1 - 2 \neff/A}
\end{equation}
}

\edit1{
This means that impact of the additional covariances among the parameters causes the uncertainties to increase such that
\begin{equation}
\frac{\sigma_f^2}{\sigma^2} = \frac{(\fisher^{-1})_{ff}}{\sigma^2} \approx \frac{8\pi a_1a_2}{1 - \frac{8\pi a_1a_2}{A}} = \boxed{\frac{2\neff}{1 - \frac{2\neff}{A}}}
\end{equation}
essentially doubling the effective PSF area. This, along with the additional three free parameters, leads to a bias of
\begin{equation}
\boxed{
\frac{\delta_{f_\ml}}{f_\ml} = \frac{5}{2} \frac{8\pi a_{1,\ml} a_{2,\ml}}{1 - \frac{8\pi a_{1,\ml} a_{2,\ml}}{A}}\frac{\sigma^2}{f_\ml^2}
}
\end{equation}
}

\end{document}